\documentclass[fleqn,usenatbib]{mnras}

\usepackage[T1]{fontenc}
\usepackage{ae,aecompl}

\usepackage{graphicx}	
\usepackage{amsmath}	
\usepackage{amssymb}	
\usepackage{xspace}
\usepackage{bm}

\newcommand{\var}{\mathop{\rm Var}\nolimits}
\newcommand{\cov}{\mathop{\rm Cov}\nolimits}

\newcommand{\expect}{\mathbb{E}}
\newcommand{\disp}{\mathbb{D}}
\newcommand{\ac}{Astron. \& Comput.\xspace}

\title[Impact of photosphere on transit timings]{The impact of photospheric
brightness field on exoplanetary transit timings and the TTV excess of HD~189733~\emph{b}}
\author[R.V.~Baluev]{Roman V. Baluev\thanks{E-mail: r.baluev@spbu.ru}\\
Saint Petersburg State University, 7--9 Universitetskaya Emb., Saint Petersburg 199034, Russia}

\begin{document}

\date{Accepted 2021 October 20.
      Received 2021 October 18;
      in original form 2021 September 17}

\pagerange{\pageref{firstpage}--\pageref{lastpage}} \pubyear{2021}

\maketitle

\label{firstpage}

\begin{abstract}
We consider the issue of excessive TTV noise observed for the exoplanet HD~189733~\emph{b}.
Trying to explain it through the host star photospheric activity, we model the stellar
surface brightness as a random field, then characterize statistical properties of the
resulting transit signal perturbation and compute individual corrections to transit timings
uncertainties. We find that possible effect of the photospheric brightness field can
explain only a minor portion ($\sim 10$~s) of the observed ($\sim 70$~s) TTV excess of
HD~189733, suggesting that the rest should be attributed to other sources. Regarding the
photospheric pattern, we place an upper limit of $\sim 0.01$ on the combination
$\varepsilon_{\rm cell} r_{\rm cell}$, where $\varepsilon_{\rm cell}$ is the relative
magnitude of brightness variations, and $r_{\rm cell}$ is the geometric cellularity scale
(relative to star radius).
\end{abstract}


\section{Introduction}
Transit timing variations, or TTVs, appear if the Keplerian motion of a transiting
exoplanet is somehow perturbed. The sources of such a perturbation can be different. For
example, it can come from another body in the system, so the TTVs can be considered as an
additional method of exoplanet detection \citep{HolmanMurray05,Agol05}. Previously, this
approach allowed to detect a statistically sound list of exoplanets from the NASA
\emph{Kepler} mission \citep{Ford11,Steffen13,Xie13}.

Alternatively, long-term TTV trends may appear because of a tidal interaction between the
planet and its host star, so that planet can exhibit either a slow spiral falling on the
star or a long-period apsidal drift. Such an example is provided by the planet WASP-12
\citep{Maciejewski16,Maciejewski18a,BaileyGoodman19}. Yet another example of this kind,
though less certain, is provided by WASP-4 \citep{Bouma19,Baluev20}. But later it appeared
that its TTV trend can be due to a long-term secular acceleration revealed in radial
velocity \citep{Bouma20}.

The present work is devoted, in general, to further advance of the TTV method, and, in
particular, to further development of the EXPANSION project (EXoPlanetary trANsit Search
with an International Observational Network). EXPANSION is a ground-based TTV project,
grown on the basis of the Exoplanet Transit Database
\citep{Baluev15a,Sokov18,Baluev19,Baluev20,Baluevetal21}. It involves an international network
of several dozens of rather small-aperture amateur and professional telescopes, aimed to
monitor exoplanetary transits. The network comprises observatories that are spreaded over
the world in the both hemispheres.

\citet{Baluev15a,Baluev19} presented TTV analysis of large sets of exoplanetary transit
lightcurves from the EXPANSION project. In these works an important issue was revealed
concerning the accuracy of the derived timing measurements. While some targets demonstrated
good agreement of their TTV noise with the estimated timing uncertainties, others revealed
an obvious TTV excess. However, that TTV excess could not be explained through a simple
deterministic model (like a TTV trend or a periodic variation). It looked like a random
noise, though with a considerably larger magnitude than one should expect from the derived
uncertainties. It does not seem that there is a clear separation between TTV-stable and
TTV-noisy stars, but the most prominent example of the second type is HD~189733~\emph{b}.
Its TTV noise is roughly twice as large as expected, and in absolute magnitude the TTV
excess is above $1$~min. This appears paradoxical, because for HD~189733 many accurate
high-quality lightcurves are available, for which such level of timing noise is well above
reasonable fitting errors. \citet{Baluev19} hypothesized that TTV excesses observed for
HD~189733 and some other EXPANSION targets might be related to stellar activity.

HD~189733 is indeed known for a strong activity, as confirmed by multiple observation
campaigns \citep{Boisse09,Cauley17,Pillitteri14}. Therefore, this star suits well as a test
case to deeply investigate the TTV noise excess and its possible cause.

Star spots and star faculae are the most obvious activity manifestation capable to distort
apparent exoplanetary transit times. See e.g. \citep{Montalto14}, and an expanded review of
the issue in \citep{Baluevetal21}. The latter work presents an attempt of a systematic search
of all statistically detectable spot-crossing events in a big photometric data base.
Although about a hundred of such potential events were detected, this comprised just a few
per cent of all transit lightcurves analysed. No obvious correlation was revealed between
the target TTV excess and the number of detected spot- and facula-crossing events.
Therefore, they cannot offer an easy explanation of the issue.

Although explicitly \emph{fittable} spots/faculae events are rare, there may be a larger
number of smaller spots or activity regions in the star photosphere that cannot be fitted
individually, or even detected, but can excite a significant perturbation to the transit
signal. Yet another phenomenon with a similar effect is photospheric granulation that can
impose a significant distortion on the fitted transit parameters \citep{Chiavassa17}. In
both these cases the transit curve perturbation acts as if the photometry involved an
additional noisy signal. Such a perturbation cannot be predicted for an individual transit,
but it can be treated statistically, as a random process.

Of course, all contemporary transit fitting techniques and software involve a dedicated and
adaptive model of the photometric noise. In particular, they should routinely take into
acount an adjustable noise level and possibly correlated (red) noise. However, such methods
are always model-dependent at some level, and there is no guarantee that we adequately fit
all types of the activity-induced photometric signals. Hence, some unforeseen biases may
appear in the timings, and this can be precisely that effect revealed in HD~189733 and
other TTV-noisy targets.

In this work we undertake an attempt to test this hypothesis with HD~189733 transit data.
Since there are multiple physical phenomena that can excite photospheric structures to be
obscured by a transiting planet, we avoid assuming any particular physical mechanism for
them. Instead of that, we consider the star photosphere as a more or less general random
field. Then we try to derive what statistical characteristics of this random field should
affect the best fitting transit parameters. This formal view would be somewhat similar to
\citet{Chiavassa17}, but with a special emphasis on the transit timings rather than depth.
Our goals are to (i) verify whether this approach can explain the TTV excess of HD~189733
and in what part, (ii) derive some numeric statistical characteristics of the stellar
photosphere as implied by such treatment or, at least, place relevant numeric limits, (iii)
develop more or less practical methods and models that could be used in future work based
on this treatment.

The paper therefore assumes the following scheme:
\begin{enumerate}
\item First of all, we give a more detailed discussion of the TTV noise excess in HD~189733
(Sect.~\ref{sec_ttvjit}).
\item We specify the statistical model of the Photospheric Brightness Field (PBF)
(Sect.~\ref{sec_bfield}).
\item We determine statistical properties of a photometric transit perturbation generated by
the PBF (Sect.~\ref{sec_pert}).
\item We explain the resulting statistical effect on the best fitting parameters of the
transit, in particular on the fitted midtimes (Sect.~\ref{sec_ttvpert}).
\item Given this TTV noise model, we solve the inverse problem, i.e. based on the
transit times available for HD~189733 try to determine statistical properties of the PBF
necessary to explain such observed timings (Sect.~\ref{sec_HD189733}).
\end{enumerate}

\section{The issue of TTV excess in HD~189733}
\label{sec_ttvjit}
\citet{Baluev19} presented a homogeneous analysis of $109$ transit lightcurves for
HD~189733, involving professional as well as amateur observations from the EXPANSION
network. It was noticed that derived transit timing measuments demonstrate an unusually
high level of TTV noise than expected from the uncertainties. First of all, let us provide
some additional details of this phenomenon.

To assess the effect of TTV noise excess, we computed a maximum-likelihood fit of these
$109$ timings using linear model for the timings $\tau_i$. We used two alternative TTV
noise models that treat transit timing uncertainies $\sigma_{\tau,i}$ differently. The
first model is the so-called multiplicative one, in which noise variance expressed as $\disp
\tau_i = \kappa^2 \sigma_{\tau,i}^2$, with fittable $\kappa$. This model corresponds to the
classic least-square fit, when the free scale factor $\kappa$ is often implicit and has the
meaning of the reduced $\chi^2$ of the residuals. The second model is the so-called
additive one, with $\disp \tau_i = \sigma_{\rm jit}^2 + \sigma_{\tau,i}^2$, where
$\sigma_{\rm jit}$ is a fittable `jitter' parameter. More details on these formal models
and their discussion are given in \citep{Baluev14a}. The maximum-likelihood fitting method
was basically the same as in \citep{Baluev08b}

\begin{table*}
\caption{Comparative properties of several TTV fits for HD~189733.}
\label{tab_ttvfits}
\begin{tabular}{lllll}
\hline
TTV noise model & Goodness-of-fit $\tilde l$ [s]$^1$ & noise factor $\kappa$, equiv. of reduced $\sqrt{\chi_\tau^2}$ & TTV jitter $\sigma_{\rm jit}$ [s] & Periodogram peak power \\
\hline
\multicolumn{5}{c}{All $109$ transit timings from \citep{Baluev19}:}\\
Multiplicative & $92.14$  & $2.11\pm 0.15$             & $0$        & $\mathbf{17.3}$ \\
Additive       & $87.96$  & $\simeq 1$ by construction & $67.8\pm 6.8$& $11.0$ \\
\multicolumn{5}{c}{Selected $42$ lightcurves, reprocessed in this work:}\\
Multiplicative & $101.15$ & $2.69\pm 0.29$             & $0$        & $\mathbf{24.1}$ \\
Additive       & $102.66$ & $\simeq 1$ by construction & $82\pm 12$ & $\mathbf{17.0}$ \\
\multicolumn{5}{c}{$109$ minus five \citet{Kasper19} lightcurves of 2016-Aug-02:}\\
Multiplicative & $74.32$  & $1.67\pm 0.12$             & $0$        & $8.9$ \\
Additive       & $79.38$  & $\simeq 1$ by construction & $53.1\pm 6.2$ & $7.5$ \\
\multicolumn{5}{c}{$42$ minus five \citet{Kasper19} lightcurves of 2016-Aug-02:}\\
Multiplicative & $59.88$  & $1.54\pm 0.18$             & $0$        & $4.6$ \\
Additive       & $75.35$  & $\simeq 1$ by construction & $23.2\pm 6.0$ & $\mathbf{12.0}$ \\
\hline
\end{tabular}
\small
$^1$ A dimensional goodness-of-fit measure directly related to the likelihood, see \citep{Baluev08b}.
\end{table*}

The results of these two TTV fits are given in Table~\ref{tab_ttvfits} (its first block).
We can see that the noise scale factor exceeds $2$, and this can be alternatively explained
through a TTV jitter of $\sim 70$~s.

\begin{figure}
\includegraphics[width=84mm]{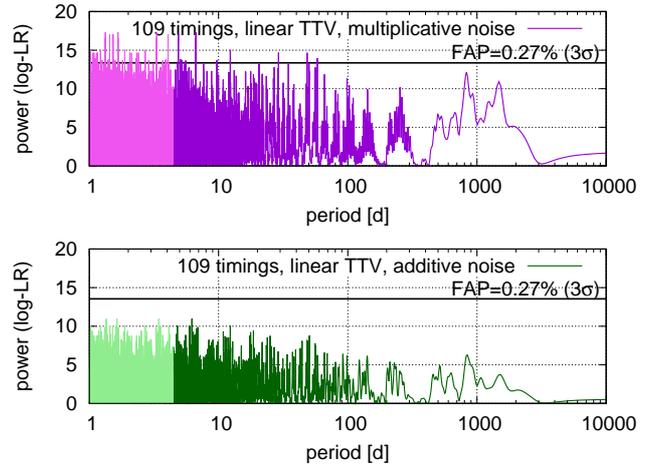}
\caption{Likelihood-based periodograms of $109$ HD~189733 timings from \citep{Baluev19},
for the first two fits from Table~\ref{tab_ttvfits}. The ordinate is the logarithm of the
likelihood ratio for best fits with and without sinusoidal signal. The significance
threshold is derived using \citep{Baluev08a} treatment. The Nyquist range is shown, in each
graph, with a darker color. The light-color range is a replica of this primary one, based
on periodogram periodic properties.}
\label{fig_prdg}
\end{figure}

Can this TTV noise excess be explained through deterministic variation? As in
\citep{Baluev19}, a possible quadratic TTV trend does not appear statistically significant
with any noise model, as it cannot provide a remarkable likelihood improvement. However,
the residual periodogram (using linear TTV model as the base) appears puzzling in the case
of multiplicative model (Fig.~\ref{fig_prdg}). We can see multiple peaks with formal
significance above the three-sigma level, but the periodogram still looks more like noise,
because there are no clearly dominating isolated periods. Therefore, the excessive TTV
variance cannot be convincingly explained through a simple regular variation, looking more
like an additinal random component.

For HD~189733 multiple amateur transit lightcurves analysed in \citep{Baluev19} involved
uncertainties or even errors regarding their metadata (e.g. whether the BJD correction was
applied or not), or about spectral bandpass (e.g. amateur filters are not always obviously
related with the standard astrophysical photometric systems). Although all such
uncertainties were apparently resolved in \citep{Baluev19}, in this work we aimed to
perform a possibly cleanest analysis. So in our primary analysis we decided not to rely on
the EXPANSION network at all, using only photometric data published in the literature:
\citet{Bakos06}, \citet{Winn07a}\footnote{Notice that published T10APT photometry revealed
double HJD correction, and we used correct data kindly provided by the authors.}, the HST
photometry acquired by \citet{Pont07} and \citet{McCullough14}, and high-accuracy broadband
photometry by \citet{Kasper19}. This builded up $42$ transit lightcurves in total.

We applied essentially the same analysis pipeline to them as in \citep{Baluev19}, which is
based on the \texttt{transitfit} command of our {\sc PlanetPack} software
\citep{Baluev13c,Baluev18c}. Namely, we fit all lightcurves using the same values for
planet/star radii ratio, transit duration, impact parameter. Only the midtimes were allowed
to vary separately. Also, the lightcurves models included individual cubic trends, and
quadratic limb darkening law with coefficients depending on the bandpass.

In this work we fit the limb darkening coefficients only for the HST and T10APT data. The
other lightcurves refer to standard photometric filters, so we used theoretically predicted
values based on \citet{ClaretBloemen11} FCM tables, taking $T_{\rm eff}=5109K$, $\log
g=4.69$, $[Fe/H]=0.03$ from \citet{Santos13}. These theoretic coefficients are listed in
Table~\ref{tab_ld}, and they appear close to those used in \citep{Kasper19}. Notice that we
additionally corrected these values by a small systematic bias derived in
\citep{Baluev19}.

\begin{table}
\caption{Limb darkening coefficients of HD~189733, for different bandpasses.}
\label{tab_ld}
\begin{tabular}{lrr}
\hline
Band   &  $A$      &  $B$ \\
\multicolumn{3}{c}{Theoretic$^1$ from \citep{ClaretBloemen11}:}\\
$B$    &  $0.839$  &  $0.004$ \\
$V$    &  $0.636$  &  $0.133$ \\
$u$    &  $1.050$  & $-0.198$ \\
$g$    &  $0.763$  &  $0.060$ \\
$r$ (same set for $R_C$)      &  $0.547$  &  $0.175$ \\
$i$ (same set for $I$, $I_C$) &  $0.436$  &  $0.192$ \\
$z$    &  $0.361$  &  $0.202$ \\
T10APT $b+y$ (not used$^2$)  &  $0.712$  &  $0.103$ \\
\multicolumn{3}{c}{Fitted in this work:}\\
HST ACS       &  $0.730\pm 0.034$  &  $-0.228\pm 0.047$ \\
HST WFC3      &  $0.326\pm 0.094$  &  $-0.04\pm 0.12$ \\
T10APT $b+y$  &  $0.77\pm 0.16$  &  $-0.11\pm 0.20$ \\
\hline
\end{tabular}\\
$^1$To these theoretic values we further added empiric corrections $\Delta A=0.004$ and $\Delta
B=-0.099$ from \citep{Baluev19}.\\
$^2$Computed simply as half sum for the $b$ and $y$ bands, but this might appear inaccurate.
\end{table}

The pipeline also involved the red noise detection and fitting algorithm described in
details in \citep{Baluev19}.

The spot-crossing anomalies detection algorithm from \citep{Baluevetal21} was not applied here,
because after removal of the EXPANSION lightcurves only the HST photometry was left with
formally significant anomalies, but they looked, in turn, like some detrending inaccuracies
rather than physical spot-crossing events (see graphs in \citealt{Baluevetal21}). Besides, one
of the HST transits turns ill fitted with such a model, if the transit midtimes are allowed
to vary.

\begin{figure*}
\includegraphics[width=\linewidth]{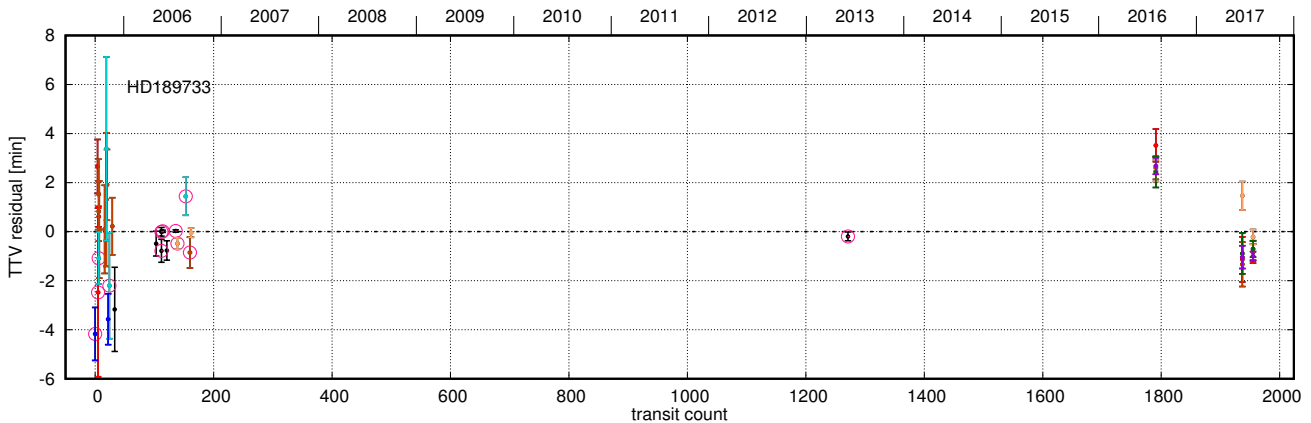}
\caption{Transit timing residuals obtained for $42$ lightcurves of HD~189733 included in
the analysis. The color of a point identifies the photometric spectal band (same convention
as in \citealt{Baluev19}). Encircled points are those a correlated photometric noise was
revealed.}
\label{fig_ttv}
\end{figure*}

The resulting set of transit times is shown in Fig.~\ref{fig_ttv}. We can see that some
points significantly deviate from the linear TTV ephemeris, in particular the group of five
timings in 2016 that belong to \citet{Kasper19}. They actually refer to a single transit
observed in five filters simultaneously.

The fits of these selected $42$ timings is given in the second block of
Table~\ref{tab_ttvfits}. We can see that the TTV noise excess remains there and is even
increased. Now both TTV models generate periodograms with unusually boosted noise level
without clear dominating peak. This indicates that neither of two models provides a
satisfactory statistical explanaition of the TTV noise.

We noticed previously that such periodogram behaviour may appear for data with outliers. In
view of that we tried to detect possibly odd lightcurves using an analogue of the
leave-one-out method. We sequentially removed a single point from our $42$ timings, then
recomputed the periodogram for this reduced set, and looked how much the periodogram global
maximum is changed. As could be expected from Fig.~\ref{fig_ttv}, the periodogram is
sensitive to the transit observed by \citep{Kasper19} on 2016-Aug-02. By removing these
five lightcurves, the TTV fits are significantly improved and the maximum periodogram power
falls below reasonable significance thresholds.

However, such treatment remains unsatisfactory. As seen from Table~\ref{tab_ttvfits} (third
and fourth blocks), the TTV noise excess is reduced indeed, but it still persists and
remains significant, so the issue is not resolved completely. On the other hand, there is
no reasonable explanation why the high-quality \citet{Kasper19} observations could involve
that large errors. Also, two other transits from \citet{Kasper19} do not indicate so big
timing deviation, even though one of them had only partial coverage.

Possible pipeline imperfections may result in systematically undervalued uncertainties, but
this effect cannot provide a convincing explanation of the issue. For example, after
per-transit averaging our timing estimates are in good agreement with those given by
\citet{Kasper19}. Besides, \citet{Baluev19} verified their timing uncertainties for
WASP-12, by comparing them with those obtained by \citep{Maciejewski16,Maciejewski18a} from
the same lightcurves but using alternative analysis. It appeared that \citet{Baluev19}
timing uncertainties may be possibly undervalued by the factor of $1.22$. It is difficult
to perform a similar comparison for HD~189733, because its data are not so homogeneous as
for WASP-12, but we can expect a similar pipeline-related effect in its $\tau_i$
uncertainties as well. Clearly, this is much smaller than the observed TTV effect. Also,
there are multiple targets in \citep{Baluev19} that demonstrate a good agreement of their
TTV noise with the derived timing uncertainties. If the issue was because of pipeline
biases, it would affect either all targets at once or primarily those with poor-quality
data (not HD~189733).

Therefore, we should deal with another explanation of the issue, namely that the observed
TTV noise of HD~189733 is owed to some physical effect in the star-planet system. Then we
should not remove the deviating \citet{Kasper19} transit, because it may carry the most
important information about the effect. The magnitude of the extra TTV noise remains then
$\sim 80$~s, corresponding to $\sim 15\%$ of the planet radius. In this work we try to
explain this TTV noise through possible photospheric brightness variations of unspecified
nature.

Our two TTV noise models appear statistically similar in terms of their goodness-of-fit,
for example the Vuong test \citep{Vuong89,Baluev12} indicates, at most, an $1.8$-sigma
significance for their difference. The additive noise model does not appear clearly better,
so our effect in question does not necessarily behave as a constant additional noise. For
example, it may affect different timings differently, so we need a nontrivial statistical
model for this putative effect.

\section{Model of photospheric brightness field}
\label{sec_bfield}
Let the brightness of star photosphere at the sky-projected position $\bmath x$ be
$I(\bmath x)$, with $x=0$ at the projected star center. We treat this $I(\bmath x)$ as a
random field characterized by some correlation function. Geometrically the photosphere is a
sphere, and in the general case there is a projection effect that makes the correlation
function (i) spatially nonstationary, and (ii) anysotropic. However, near the star center
we can neglect the projection effect and consider the local correlation function $k_I$:
\begin{align}
k_I(\bmath x-\bmath x') &= \cov(I(\bmath x),I(\bmath x')) = \expect(\delta I(\bmath x) \delta I(\bmath x')), \nonumber\\
\delta I(\bmath x) &= I(\bmath x) - \expect I(\bmath x), \quad x,x'\ll R_\star.
\label{kI}
\end{align}
We assume that $k_I(\bmath r)$ in~(\ref{kI}) is stationary (shift-invariable) and so it
depends on just a single argument. We also assume that it is radially symmetric, i.e. it
can be rewritten as $k_I(r)$ with a scalar argument $r$. The spherical geometry distorts
both these properties in a mathematically predictable way, but at the local level we may
start from such a $k_I$. In this work we do not address possible effects of an intrinsic
(non-projectional) anysotropy of the brightness field, and also we do not consider possible
temporal variations in $I(\bmath x)$.

Based on the Wiener-Khinchin theorem, we can construct the two-dimensional spatial power
spectrum $P_I(s)$, which is also radially symmetric:
\begin{equation}
P_I(s) = \frac{1}{4\pi^2} \int\limits_{\mathbb R^2} k_I(\bmath r) e^{i \bmath s \bmath r} d\bmath r =
 \frac{1}{2\pi} \int\limits_0^{+\infty} k_I(r) J_0(sr) r dr.
\label{powI}
\end{equation}
The total power is then
\begin{equation}
k_I(0) = \int\limits_{\mathbb R^2} P_I(\bmath s) d\bmath s,
\end{equation}
which appears equal to the variance $\var I(\bmath x)$ and is constant in stationary
approximation.

Notice that $I(\bmath x)$ may often have a \emph{cellular} structure, i.e. it may remain
nearly constant within small geometric units, or cells, in which the correlation keeps
high. Therefore, $k_I(0)$ has the meaning of spatial brightness variation between such
independent cells. To highlight this, we define $\sigma_{\rm cell}^2=k_I(0)$.

In addition to $k_I(0)$, an $x$-space parameter, let us consider $P_I(0)$, being an
analogous $s$-space characteristic:
\begin{equation}
P_I(0) = \frac{1}{4\pi^2}\int\limits_{\mathbb R^2} k_I(\bmath r) d\bmath r.
\end{equation}
From physical dimensionality of $P_I(0)$ it is tentative to introduce a quantity $r_{\rm
cell}$ that has the meaning of a spatial scale, based on the following definition:
\begin{equation}
\sigma_{\rm cell}^2 r_{\rm cell}^2 = 4\pi^2 P_I(0) = \int\limits_{\mathbb R^2} k_I(\bmath r) d\bmath r =
2\pi \int\limits_0^{+\infty} k_I(r) r dr.
\end{equation}
That is, $r_{\rm cell} = 2\pi \sqrt{P_I(0)/k_I(0)}$. This $r_{\rm cell}$ will appear as one
of key quantities below, so let us try to understand its physical meaning. Often it may
serve as a measure of a cellularity scale in the random field, i.e. the typical size of a
single correlation cell. However, counterexamples exist, in which the cellularity scale is
different from $r_{\rm cell}$. Let us now consider a few simple demonstrative cases.
\begin{enumerate}
\item \emph{White-noise field.} Let $P_I(s)$ is constant for $s<r_0^{-1}$
with some small $r_0$, and zero otherwise. Then $k_I(r) = 2 \sigma_{\rm cell}^2 J_1(r/r_0)
r_0/r$ and $r_{\rm cell} = 2 r_0\sqrt\pi$. In this case $k_I(r_{\rm cell})/k_I(0)\approx
0.07$, meaning that brightness at two points separated by $r_{\rm cell}$ appears nearly
uncorrelated, and hence such points likely belong to independent cells.

\item \emph{Red-noise field.} Let $k_I(r) = \sigma_{\rm cell}^2 e^{-r^2/(2r_0^2)}$.
Then $P_I(s) = \sigma_{\rm cell}^2 r_0^2 e^{-r_0^2 s^2/2}/(2\pi)$ and $r_{\rm cell} = r_0
\sqrt{2\pi}$. In this case $k_I(r_{\rm cell})/k_I(0)\approx 0.04$, again a small
correlation similar to the one from the first case.

\item \emph{Blue-noise field.} Let $P_I(s) \propto s^2 e^{-r_0^2 s^2/2}$. Contrary to
previous cases, this power spectrum is peaked at a non-zero $s$, close to $r_0^{-1}$, while
$P_I(0)=0$. The extra factor $s^2$ in this definition can be obtained by applying the
Laplace operator to $k_I$ from the previous example, so after proper normalization it
should be $k_I(r) = \sigma_{\rm cell}^2 \kappa(r/r_0)$, where $\kappa(u) = (1-u^2/2)
e^{-u^2/2}$. This example has $r_{\rm cell}=0$ for any $r_0$, so $r_{\rm cell}$ becomes
non-physical. A reasonable measure of cellularity scale can be set to the same value as in
the previous example, $r_{\rm cell}' = r_0 \sqrt{2\pi}$.
\end{enumerate}

As we can see, $r_{\rm cell}$ describes the cellularity scale well if $k_I(r)\geq 0$
everywhere, but issues may appear if $k_I$ is sign-changing. An alternative measure could
be based on integrating $|k_I|$ or $k_I^2$. But, for example, in the white-noise case the
integral of $|k_I|$ is infinite, while the integral of $k_I^2$ yields exactly the same
value, $r_{\rm cell}=2r_0\sqrt\pi$. Yet another way is to define the cellularity scale
through the first zero of $k_I(r)$, which is $r=1.08 r_{\rm cell}$ for the white noise, but
this definition does not work if $k_I\geq 0$. In general, it appears not so easy to
construct a universally good measure of cellularity scale, but our $r_{\rm cell}$ often
retains this meaning (and even for sign-changing $k_I$ cases).

\begin{figure*}
\includegraphics[width=\linewidth]{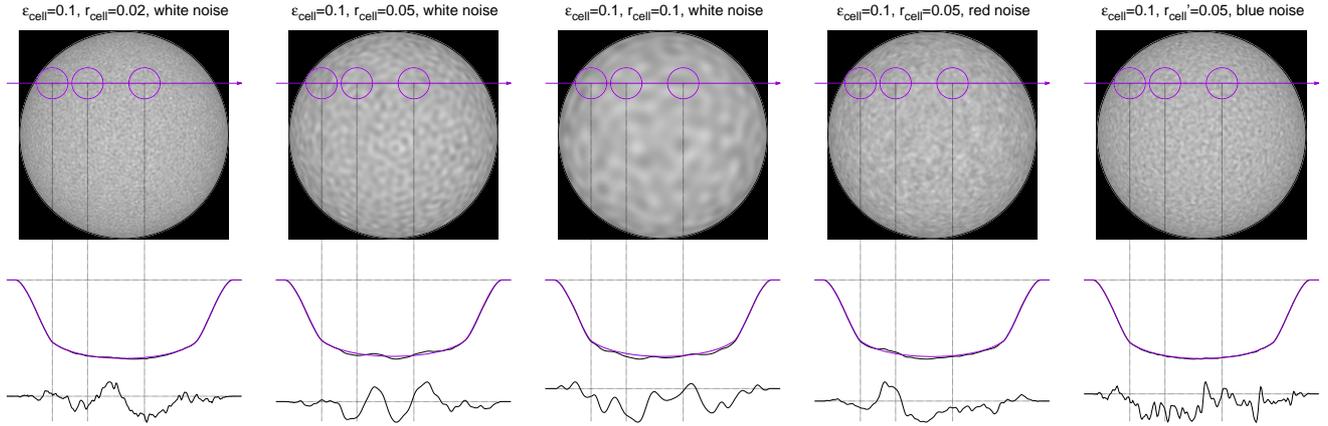}
\caption{Modelling the stellar PBF: the simulated photosphere image (top row), the
resulting transit signal (middle row), and the implied perturbation magnified (bottom row).
The transiting planet has the radius of $0.15 R_\star$, the transit impact parameter is
$b=0.5$, and the limb darkening coefficients are $A=0.4$ and $B=0.2$. See text for more
details.}
\label{fig_BF}
\end{figure*}

Several simulated examples of our toy models are shown in Fig.~\ref{fig_BF}. These random
fields were generated on a grid of $120\times 120$ pixels and take into account the
spherical curvature of stellar surface. This was achieved by setting the argument of
$k_I(r)$ to $r = 2\sin\frac{\alpha}{2}$, where $\alpha$ is the angle between two 3D unit
vectors $\bmath w$ and $\bmath w'$ determining positions on a sphere with $R_\star=1$. The
components $w_x$ and $w_y$ represented image coordinates, while $w_z$ was derived as $w_z =
\sqrt{1-w_x^2-w_y^2}$. In such a way we constructed the full covariance matrix for all
$14400$ image points, and then simulated the PBF image as a multivariate Gaussian
vector.\footnote{The transition from Cartesian to spherical random field is unobvious in
what concerns large-scale correlations. A more straightforward choice would be to set $r$
proportional to the arc $\alpha$, but this frequently generated a nonphysical covariance
matrix (not strictly positive definite). The chord-based version, $r \propto
\sin\frac{\alpha}{2}$, empirically allowed to bypass this issue. An entirely strict
approach should likely involve spherical harmonics, but this is unnecessary for our sample
simulations.}

The images were also post-modulated by a quadratic limb-darkening law ($A=0.4,B=0.2$). All
plots correspond to the same value of the relative quantity $\varepsilon_{\rm cell} =
\sigma_{\rm cell}/I_0 = 0.1$, where $I_0$ is the nominal brightness at the disc center. The
value of $r_{\rm cell}$ is labelled in each plot (it is assumed relative to $R_\star$).

For each simulated PBF we then computed (numerically) a transit signal for a potential
planet with $r_{\rm pl} = R_{\rm pl}/R_\star=0.15$. These transit curves are shown in the
bottom part of Fig.~\ref{fig_BF}. We can see that they reveal a randomly varying
perturbation, owed to photospheric inhomogeneities. Our further goal is to determine the
statistical properties of this perturbation, based on the $k_I(r)$.

In case of the Sun, the spatial power spectrum of its photospheric brightness field was
investigated decades ago, see e.g. \citet{Karpinskii77}. In terms of our $P_I(s)$, this
spectrum remains more or less constant down to the granulation scale $\sim 1000$~km, where
it starts to quickly decrease. Therefore, solar surface should better correspond to the
white-noise example above (though its granularity is more fine than shown in any PBF of
Fig.~\ref{fig_BF}). However, this likely refers to only inactive photospheric domains, the
effect of multiple spots on $P_I$ is not very clear.

Notice that we do not assume here any particular correlation function $k_I$ or the power
spectrum $P_I(s)$, as our approach will not require to specify them in full. The simulated
images in Fig.~\ref{fig_BF} are just examples, and we can construct different patterns by
augmenting $P_I(s)$ with a more complicated behaviour. For example, it is possible to
control the internal structure of a cell. Granulation pattern would require cells with a
wider core and thin boundaries, while a spotted pattern may need smaller cell cores
separated by relatively wider space. Such fine-tuning would require a more clever
construction of $P_I(s)$ in the high-$s$ range, but should not affect its low-$s$
behaviour.

\section{Characterizing the transit perturbation signal}
\label{sec_pert}
Based on the PBF perturbation $\delta I$, let us write down the subplanet flux perturbation
as
\begin{equation}
\delta F(\bmath x) = \int p(\bmath x' - \bmath x) \delta I(\bmath x') d\bmath x',
\end{equation}
where $p(\bmath r)$ is the indicator function of the projected planetary disc (unit for
$r<r_{\rm pl}$, and zero otherwise), $\bmath x$ is the position of planet center. As
before, we assume $R_\star=1$.

To understand the issue better, let us first consider a simplified case when $x$ and $x'$
are small (near the star disc centre), and therefore $k_I$ is shift-invariable. Then the
correlation function of $\delta F$ is also shift-invariable and it can be expressed through
a convolution:
\begin{align}
k_F(\bmath x-\bmath x') &= \expect(\delta F(\bmath x) \delta F(\bmath x')) \nonumber\\
&= \iint p(\bmath x'' - \bmath x) p(\bmath x''' - \bmath x') k_I(\bmath x'' - \bmath x''') d\bmath x'' d\bmath x''' \nonumber\\
&= \iint p(\bmath x'' - \bmath x) p(\bmath x'' - \bmath x' - \bmath r) k_I(\bmath r) d\bmath x'' d\bmath r \nonumber\\
&= \int k_p(\bmath r + \bmath x' - \bmath x) k_I(\bmath r) d\bmath r = (k_I * k_p)(\bmath x-\bmath x'), \nonumber\\
k_p(\bmath r) &= \int p(\bmath x) p(\bmath x - \bmath r) d\bmath x = (p * p)(\bmath r).
\label{kF}
\end{align}
Therefore, $k_F(r)$ can be viewed as the ``geometric kernel'' $k_p(r)$, blurred
by $k_I(r)$, and all functions appear radially symmetric.

Notice that $k_p(r)$ equals to the intersection area beneath two circles, with radii
$r_{\rm pl}$ both and with centers separated by $r$. Simple geometric constructions yield
\begin{equation}
k_p(r) = r_{\rm pl}^2\, \kappa\left(\frac{r}{r_{\rm pl}}\right), \quad \kappa(u) = 2\arccos \frac{u}{2} - u \sqrt{1-\frac{u^2}{4}}.
\end{equation}

We do not aim to adopt any specific $k_I$ here. However, we need to make a
no-so-restrictive assumption that $k_I$ has a narrow localization, much smaller than
$r_{\rm pl}$. In this case, the convolution $k_p * k_I$ should impose only a negligible
smoothing effect on $k_p$, regardless of the particular shape of $k_I$. Then slowly varying
$k_p$ can be moved out of the integration in~(\ref{kF}):
\begin{equation}
k_F(r) \simeq k_p(r) \int k_I(\bmath r') d\bmath r' = k_p(r) \sigma_{\rm cell}^2 r_{\rm cell}^2.
\label{kca}
\end{equation}
Therefore, the effect of $k_I$ is approximately equivalent to multiplying by a constant.

Now let us consider this effect in the Fourier space. The spatial power spectra obey, in
turn, the multiplication low:
\begin{equation}
P_F(s) = P_I(s) P_p(s).
\label{Pca}
\end{equation}
According to our assumption, $k_I$ is narrow localized relative to $k_p$, hence the power
spectra $P_I$ and $P_p$ should obey the opposite relationship. Therefore, regardless of a
particular shape of $P_I$, we can replace $P_I(s)$ by $P_I(0)$ in~(\ref{Pca}), and this
leads us to an equivalent multiplication by a constant.

\begin{figure}
\includegraphics[width=84mm]{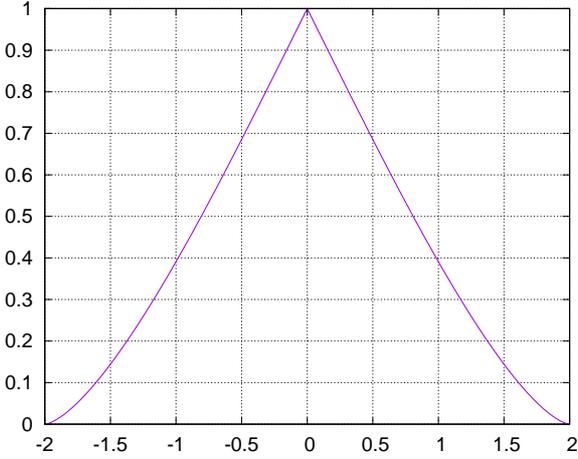}
\caption{Correlation function $k_p(r)$, in normalized axes.}
\label{fig_k_p}
\end{figure}

Therefore, neglecting the projection effects, the correlation function $k_F(r)$ should be
close to $k_p(r)$. A graph of this function is shown in Fig.~\ref{fig_k_p}. Its shape is
close to a triangle, with the localization range $[-2r_{\rm pl}, 2r_{\rm pl}]$.

Now we should extend our approach to general conditions, involving near-limb projection.
Then the PBF becomes not shift-invariable because of spherical curvature of the
photosphere, hence $k_I$ becomes a function of two arguments. We can generalize~(\ref{kF})
as follows:
\begin{align}
k_F(\bmath x,\bmath x') &= \expect(\delta F(\bmath x) \delta F(\bmath x')) \nonumber\\
&= \int p(\bmath x'' - \bmath x) p(\bmath x''' - \bmath x') k_I(\bmath x'', \bmath x''') d\bmath x'' d\bmath x'''.
\end{align}
As before, we assume that $k_I$ has narrow localization in comparison with $p$, meaning
that we may equate $x''=x'''$ everywhere in the integrand, except $k_I$ itself:
\begin{equation}
k_F(\bmath x, \bmath x') = \int p(\bmath x'' - \bmath x) p(\bmath x'' - \bmath x') d\bmath x'' \int k_I(\bmath x'', \bmath x''') d\bmath x'''.
\label{kFgen}
\end{equation}
In the shift-invariable case, the inner integral in~(\ref{kFgen}) was equal to the constant
$\sigma_{\rm cell}^2 r_{\rm cell}^2$, but now it is not so simple and may depend on $\bmath
x''$. We need to dig into the properties of $k_I$ to treat this dependence, so let us
consider some effects that affect $k_I$ and this integral.

\begin{enumerate}
\item Considering an arbitrary sky-projected position $\bmath x''$ in~(\ref{kFgen}), the
projection angle $\theta$ would be defined from $x''=\sin\theta$. This projection effect
anisotropically compresses the PBF pattern by the factor $\cos\theta$ in radial direction.
However, $\cos\theta$ may vary only negligibly inside a single correlation cell. Hence,
$k_I$ can be treated shift-invariable in the local sense. Inside its localization domain,
i.e. for $\bmath x'''$ within an $\sim r_{\rm cell}$ distance from $\bmath x''$, we may
assume that $k_I$ depends on just the difference $\bmath x''-\bmath x'''$. This dependence
becomes anisotropic though. In turn, the inner integral in~(\ref{kFgen}) should be reduced
by the factor $\cos\theta$ because of the radial scale compression. That is, the integral
should be amended to $\sigma_{\rm cell}^2 r_{\rm cell}^2 \cos\theta$, where
$\cos\theta=\sqrt{1-{x''}^2}$.

\item The limb darkening scales the apparent brightness according to a certain law,
$\expect I = I_{\rm ld}(\bmath x)$, and we assume that it scales the perturbation field
$\delta I$ analogously. As long as $k_I$ depends on $\delta I$ in a quadratic manner, the
inner integral in~(\ref{kFgen}) should involve an additional factor $I_{\rm ld}^2(\bmath
x'')/I_0^2$. Notice that we assume a quadratic limb darkening model, $I_{\rm ld}/I_0=
Q(\cos\theta) = 1 - A (1-\cos\theta) - B (1-\cos\theta)^2$.
\end{enumerate}

Combining these two conclusions together, we have
\begin{equation}
\int k_I(\bmath x'', \bmath x''') d\bmath x''' \simeq \sigma_{\rm cell}^2 r_{\rm cell}^2 Q^2(\cos\theta) \cos\theta,
\end{equation}
and~(\ref{kFgen}) turns into
\begin{equation}
k_F(\bmath x, \bmath x') \simeq \sigma_{\rm cell}^2 r_{\rm cell}^2 \int p(\bmath x'' - \bmath x) p(\bmath x'' - \bmath x') Q^2(\cos\theta) \cos\theta d\bmath x''.
\label{kFfin}
\end{equation}
This formula can be rewritten as follows:
\begin{equation}
\frac{k_F(\bmath x, \bmath x')}{F_\star^2} = \varkappa^2 k_{\rm pert}(\bmath x, \bmath x'),
\label{kFkappa}
\end{equation}
where
\begin{equation}
\varkappa = \varepsilon_{\rm cell} r_{\rm cell} = 2\pi \frac{\sqrt{P_I(0)}}{I_0},
\end{equation}
and
\begin{align}
k_{\rm pert}(\bmath x, \bmath x') &= \frac{I_0^2}{F_\star^2} \int p(\bmath x'' - \bmath x) p(\bmath x'' - \bmath x') Q^2(\cos\theta) \cos\theta d\bmath x'', \nonumber\\
\frac{F_\star}{I_0} &= \pi \left(1-\frac{A}{3}-\frac{B}{6}\right),
\label{kFpert}
\end{align}
with $F_\star$ being the full out-of-transit flux from the star.

In the left part of~(\ref{kFkappa}) we have, in fact, the correlation function of $\delta
F/F_\star$. It describes the relative flux change typically dealt with in transit fitting.
In the right part we have the adimensional $k_{\rm pert}$ function scaled by $\varkappa^2$
(also adimensional). While $k_{\rm pert}$ is computed from~(\ref{kFpert}) entirely
theoretically, the factor $\varkappa$ is a physical parameter characterizing the
photospheric cellular pattern. This is the only characteristic left from $k_I$.

The integral in~(\ref{kFpert}) contains rather simple well defined quantities, so it is
clear in principle. It can be viewed as a more complicated version of some integrals that
appear when computing the transit curve \citep{AbubGost13} and the Rossiter-McLaughlin
effect \citep{BaluevShaidulin15}. For example, the transit signal (or the relative
in-transit flux drop) can be expressed as
\begin{equation}
\frac{\Delta F(\bmath x)}{F_\star} = \frac{I_0}{F_\star} \int p(\bmath x'' - \bmath x) Q(\cos\theta) d\bmath x'',
\end{equation}
which has a similar structure as~(\ref{kFpert}). It appears that our $k_{\rm pert}$ can be
computed analogously, by expressing it through incomplete elliptic integrals.
Unfortunately, this derivation appeared huge, as well as the final result. Moreover, at
some levels it involved multiple conditional branches that are easy to implement
algorithmically, but not so convenient to write down mathematically. In the online-only
supplement we provide a C++ code that can compute $k_{\rm pert}(\bmath x, \bmath x')$.

\begin{figure}
\includegraphics[width=84mm]{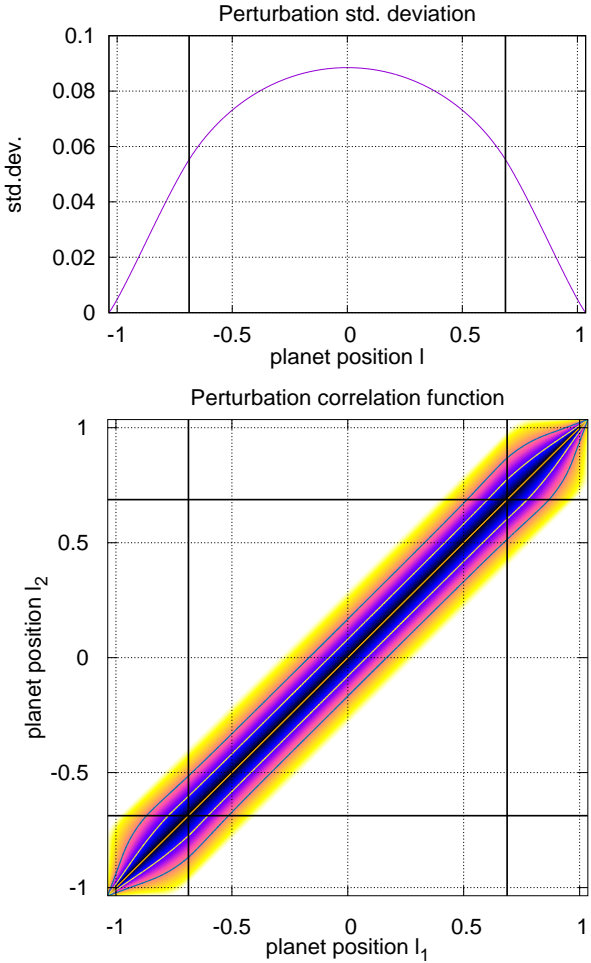}
\caption{Top: Standard deviation
$\sqrt{k_{\rm pert}(\bmath x,\bmath x)}$ as function of the transiting planet position.
Bottom: The normalized correlation function $k_{\rm pert}(\bmath x,\bmath x')/\sqrt{k_{\rm
pert}(\bmath x,\bmath x) k_{\rm pert}(\bmath x',\bmath x')}$ as function of two independent
positions of transiting planet. The planet and star parameters, and geometry of the
transit, are the same as in Fig.~\ref{fig_BF}. Quantity $l$ (or $l_{1,2}$) stand for the
planet 1D position along its transit path. Additional thick lines in the graphs label
positions of the second and third contact.}
\label{fig_pertcov}
\end{figure}

Though the derivation of $k_{\rm pert}$ appeared complicated, its graphical view is pretty
simple. It is shown in Fig.~\ref{fig_pertcov}. We can see that the variance of the
perturbation expectedly decreases when planet moves from the central transit phase to a
star limb. The correlation function in the middle of the transit appears similar to the
triangular shape shown in Fig.~\ref{fig_k_p}, but in the ingress or egress phases it
becomes more narrow.

\begin{figure}
\includegraphics[width=84mm]{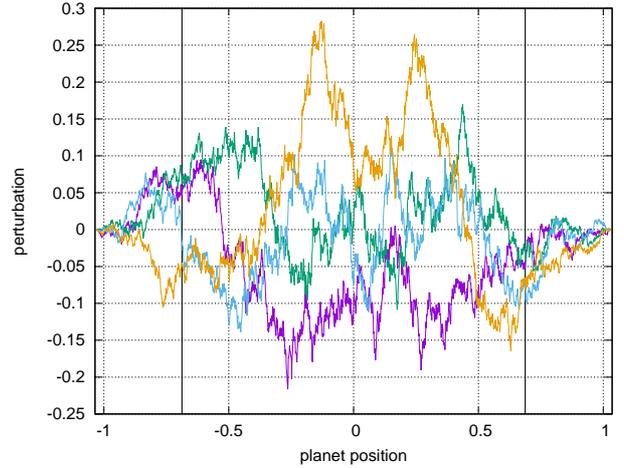}
\caption{Several simulated examples of a Gaussian process that obeys the correlation
function from Fig.~\ref{fig_pertcov}. Two additional thick lines label positions of the
second and third contact.}
\label{fig_pertsig}
\end{figure}

Also, in Fig.~\ref{fig_pertsig} we present several simulated examples of the perturbation
with correlation function from Fig.~\ref{fig_pertcov}. Comparing them with the perturbation
curves shown in Fig.~\ref{fig_BF}, we may notice that they are generally similar, except
that perturbations in Fig.~\ref{fig_BF} become more smooth for larger $r_{\rm cell}$. This
is an expected degradation of our approach accuracy, appearing if $k_I$ becomes wider
localized.

Our approximation of $k_F$ may become invalid or inaccurate in three cases: (i) if $k_I$ is
not localized well enough, for example if $r_{\rm cell}$ exceeds $r_{\rm pl}$, and (ii) if
$r_{\rm cell}$ turns zero, like in the blue-noise field considered above, and (iii) if the
PBF exhibits significant temporal variation.

The first case simply means that our approach may not work well with large-scale PBF
patterns that exceeds the planet size. In the second case, our primary approximation
term~(\ref{kca},\ref{kFfin}) vanishes because of the degeneracy, and the shape of $k_F$ is
determined by higher order terms that we neglected. However, the magnitude of photometric
perturbation should then become much smaller. Hence, PBFs of this type cannot generate
large photometric perturbation, so in the context of our analysis there is no practical
need to model them accurately. The third limitation appears if the PBF variation timescale
becomes smaller than the correlation timescale of $k_{\rm pert}$, which is equal to the
time that planet needs to pass its diameter (designate it $2\tau_{\rm pl}$). Solar
granulation has $\sim 10$~min changing timescale \citep{Nesis02}, and this appears
comparable to $2\tau_{\rm pl}$ for a typical hot Jupiter. Then the correlation wings of
$k_{\rm pert}$ should be additionally suppressed by some PBF temporal stability function,
effectively reducing the correlation timescale. However, even in such a case this effect
looks minor or moderate, at most. Moreover, Solar-type granulation has $\varkappa \lesssim
10^{-3}$ and, as we will see below, such a signal does not seem detectable in TTV data like
those we have for HD~189733, while possible PBF pattern for this target should have a much
larger scale, and hence to be more stable temporally.

\section{The PBF effect on transit fits}
\label{sec_ttvpert}
In terms of the magnitude, rather than flux, the perturbation signal is $\delta m \simeq
1.086\, \delta F/F_\star$. Therefore, the correlation function for the magnitude is $k_m
\simeq 1.086^2 k_F/F_\star^2$. From the other side, using~(\ref{kFkappa}) we have
\begin{equation}
k_m(\bmath x, \bmath x') = h k_{\rm pert}(\bmath x, \bmath x'), \quad h=(1.086\varkappa)^2,
\label{kmkappa}
\end{equation}
where the sky-projected planet positions $\bmath x$ or $\bmath x'$ depend on the
observation time $t,t'$ in a deterministic way.

For any discrete set of observations we can build up the perturbation vector $\bmath{\delta
m} = \{\delta m(t_i)\}_{i=1}^N$, and then construct its $N\times N$ covariance matrix in
the following form:
\begin{equation}
\mathbfss K_m = h \mathbfss K_{\rm pert}, \quad (\mathbfss K_{\rm pert})_{ij} = k_{\rm pert}(\bmath x(t_i), \bmath x(t_j))\, \delta_{n_i n_j},
\end{equation}
where $n_i$ and $n_j$ are ordinal transit numbers which the observations $i$ and $j$ belong
to, and $\delta_{n_i n_j}$ is their Kronecker delta (it forces zero correlation between
different transits).

Basically, the perturbation $\bmath{\delta m}$ can be viewed as an extra correlated noise
that contaminates our lightcurve. Not only it is correlated, but also nonstationary. This
perturbation will introduce biases to our lightcurve fit, in particular biases to the
photometric noise model. This would not be a big issue, if this noise model could provide
an accurate treatment for the perturbation. However, we typically use stationary noise
models in practice. In such a case primarily the central part of the transit would be
fitted well, but not the ingress/egress ranges, where the perturbation behaves differently
(e.g. it has a smaller magnitude and is more short-scale correlated). On the other hand,
timings depend more on the ingress/egress phases. Hence, a systematic distortion can be
expected in them, and our further goal is to characterize it.

Let us consider this task in general. After all, we always use some method of fitting, i.e.
a recipe how to construct the best fitting estimate $\hat\bxi$ for a vector of free
parameters $\bxi$. Also, the method should yield some matrix $\hat{\bm{\mathsf\Xi}}$ as an
estimate for the covariance matrix $\var\hat\bxi$.\footnote{Here we omit more complicated
cases when the uncertainties are asymmetric and/or the error domains are nonelliptic.}
Therefore, both $\hat\bxi$ and $\hat{\bm{\mathsf\Xi}}$ are some functions of the input data
$\bmath m$, and they would change if $\bmath m$ is replaced by $\bmath m' = \bmath m +
\bmath{\delta m}$. In the second-order approximation,
\begin{align}
\delta\hat\xi_i = \hat\xi'_i - \hat\xi_i &\simeq \sum_k \frac{\partial \hat\xi_i}{\partial m_k} \delta m_k + \frac{1}{2} \sum_{k,l}\frac{\partial^2 \hat\xi_i}{\partial m_k \partial m_l} \delta m_k \delta m_l, \nonumber\\
\delta\hat\Xi_{ij} = \hat\Xi_{ij}' - \hat\Xi_{ij} &\simeq \sum_k \frac{\partial\hat\Xi_{ij}}{\partial m_k} \delta m_k + \frac{1}{2} \sum_{k,l}\frac{\partial^2\hat\Xi_{ij}}{\partial m_k\partial m_l} \delta m_k \delta m_l.
\label{xiseries}
\end{align}
By our input conditions, $\expect\delta m_k=0$ and $\expect \delta m_k \delta m_l =
h (K_{\rm pert})_{kl}$. Therefore, the biases in $\hat\bxi$ and $\hat{\bm{\mathsf\Xi}}$ are:
\begin{align}
\expect \delta\hat\bxi &\simeq h \bxi_{\rm bias}, &(\xi_{\rm bias})_i &= \frac{1}{2} \sum_{k,l}\frac{\partial^2 \hat\xi_i}{\partial m_k \partial m_l} (K_{\rm pert})_{kl}, \nonumber\\
\expect \delta\hat{\bm{\mathsf\Xi}} &\simeq h \bm{\mathsf\Xi}_{\rm bias}, &(\Xi_{\rm bias})_{ij} &= \frac{1}{2} \sum_{k,l}\frac{\partial^2\hat\Xi_{ij}}{\partial m_k\partial m_l} (K_{\rm pert})_{kl}.
\label{pbias}
\end{align}
These biases reflect only a systematic part of a perturbation, while there is also random
part for which we can compute the covariance matrix (for $\delta\hat\bxi$), the $4$-index
covariance tensor (for $\delta\hat{\bm{\mathsf\Xi}}$), and the $3$-index tensor for the
cross-covariance between $\delta\hat\bxi$ and $\delta\hat{\bm{\mathsf\Xi}}$. All these
quantities can be approximated using only the linear terms of~(\ref{xiseries}):
\begin{align}
\var \delta\hat\bxi &\simeq h \bm{\mathsf \Xi}_2, &(\Xi_2)_{ij} &= \sum_{p,q} \frac{\partial \hat\xi_i}{\partial m_p}\frac{\partial \hat\xi_j}{\partial m_q} (K_{\rm pert})_{pq}, \nonumber\\
\cov(\delta\hat\bxi,\delta\hat{\bm{\mathsf\Xi}}) &\simeq h \mathsf \Xi_3, &(\Xi_3)_{i,kl} &= \sum_{p,q} \frac{\partial \hat\xi_i}{\partial m_p}\frac{\partial\hat\Xi_{kl}}{\partial m_q} (K_{\rm pert})_{pq}, \nonumber\\
\var \delta\hat{\bm{\mathsf\Xi}} &\simeq h \mathsf \Xi_4, &(\Xi_4)_{ij,kl} &= \sum_{p,q} \frac{\partial\hat\Xi_{ij}}{\partial m_p} \frac{\partial\hat\Xi_{kl}}{\partial m_q} (K_{\rm pert})_{pq}.
\label{pvar}
\end{align}
Perturbational characteristics defined in~(\ref{pbias}) and~(\ref{pvar}) are linear with
respect to $h$ and $\mathbfss K_{\rm pert}$. They in fact represent first terms of more
general power series in $h$, so $h$ should be small.

All these quantities can be computed analytically, though such a computation remains rather
hard for the most. Nevertherless, we derive an analytic approximation for the most simple
factor, $\bm{\mathsf\Xi}_2$, because it involves only the gradient of $\hat\bxi$. This
derivation is layed out below.

First of all, we should specify the method of fitting as different methods may result in
different quantities~(\ref{pbias},\ref{pvar}). We consider the maximum-likelihood Gaussian
process fitting following from \citep{Baluev13a} that was used in this work and in
\citep{Baluev19}. Our input data are the vector of photometric magnitudes $\bmath m$, and
they should be fitted as $\bmath m = \bmu(\btheta) + \bmath n$. Here $\bmu(\btheta)$
includes transit model and various deterministic trends, jointly parametrized by the
lightcurve parameters $\btheta$. The noise vector $\bmath n$ follows a multivariate
Gaussian distribution with the covariance matrix $\var \bmath n = \mathbfss V(\boldeta)$
that depends on another set of parameters $\boldeta$ (the noise parameters). That vector
$\boldeta$ usually includes variances of the red and white photometric noise treated as
fittable quantities, and the red noise correlation timescales. The joint vector of free
parameters is a combination of these two parametric sets, $\bxi = \{\btheta,\boldeta\}$.

The likelihood function of this task, in a classic definition, is given by
\begin{equation}
\log \mathcal L(\bxi) = -\frac{1}{2} \log \det\mathbfss V - \frac{1}{2} \bmath r^{\rm T} \mathbfss V^{-1} \bmath r, \quad \bmath r=\bmu-\bmath m.
\label{ll}
\end{equation}
This function should be maximized to obtain the best fitting estimation of $\btheta$ and
$\boldeta$. However, following \citet{Baluev08b}, we typically amend this definition to
\begin{equation}
\log \tilde{\mathcal L}(\bxi) = -\frac{1}{2} \log \det\mathbfss V - \frac{1}{2\gamma} \bmath r^{\rm T} \mathbfss V^{-1} \bmath r, \quad \gamma=1-\frac{\dim\btheta}{N}.
\label{llm}
\end{equation}
This modification allows to significantly reduce the bias in the noise parameters
$\boldeta$ that appears because the residuals $\bmath r$ represent, after the fitting, a
systematically undervalued estimation of $\bmath n$.

The best fitting estimation of $\bxi$ is given by the position of the maximum:
\begin{equation}
\hat\bxi = \arg\max \tilde{\mathcal L}(\bxi).
\label{llopt}
\end{equation}
Mathematically, the necessary condition for the maximum is that gradient of
$\log\tilde{\mathcal L}$ must vanish:
\begin{equation}
\left. \frac{\partial \log \tilde{\mathcal L}}{\partial\bxi}\right|_{\bxi=\hat\bxi}= 0.
\label{zerog}
\end{equation}
The covariance matrix of this $\hat\bxi$ is then approximated as
\begin{equation}
\hat{\bm{\mathsf\Xi}} \simeq \mathbfss F_\bxi^{-1}, \quad \mathbfss F_\bxi = \left( \begin{array}{cc} \mathbfss F_\btheta & \mathbfss 0\\ \mathbfss 0 & \mathbfss F_\boldeta \end{array}\right),
\label{estvar}
\end{equation}
where $\mathbfss F_\bxi$ is the Fisher information matrix for $\hat\bxi$. Notice that it
has a diagonal-block form with zero offdiagonal blocks, corresponding to the correlation
between $\btheta$ and $\boldeta$. Thanks to this, $\hat\btheta$ and $\hat\boldeta$ are
asymptotically uncorrelated for $N\to\infty$ (though some correlation may appear via
higher-order terms in $1/N$). The expression for the $\btheta$-part of this matrix reads:
\begin{equation}
\mathbfss F_\btheta = \mathbfss Q = \mathbfss J^{\rm T} \mathbfss V^{-1} \mathbfss J, \quad \mathbfss J=\frac{\partial\bmu}{\partial\btheta}.
\label{mQ}
\end{equation}
The expression for $\mathbfss F_\boldeta$ is not used here.

Now, let us alter the input data $\bmath m$ by adding a $\bmath{\delta m}$ perturbation.
The condition of the best fit~(\ref{zerog}) should be identically satisfied for any
$\bmath{\delta m}$. By differentiating~(\ref{zerog}) with respect to $\bmath{\delta m}$, we
obtain
\begin{eqnarray}
\frac{\partial^2 \log \tilde{\mathcal L}}{\partial\bxi\partial(\bmath{\delta m})} + \frac{\partial^2 \log \tilde{\mathcal L}}{\partial\bxi^2} \frac{\partial\hat\bxi}{\partial(\bmath{\delta m})} = 0 \implies \nonumber\\
\implies \frac{\partial\hat\bxi}{\partial(\bmath{\delta m})} = \left(\frac{\partial^2 \log \tilde{\mathcal L}}{\partial\bxi^2}\right)^{-1} \frac{\partial^2 \log \tilde{\mathcal L}}{\partial\bxi\partial(\bmath{\delta m})}.
\label{Jp}
\end{eqnarray}
Based on the definitions~(\ref{ll},\ref{llm}), we can compute the following derivatives:
\begin{equation}
\frac{\partial \log \tilde{\mathcal L}}{\partial(\bmath{\delta m})} = \frac{\bmath r^{\rm T} \mathbfss V^{-1}}{\gamma} \implies \frac{\partial^2 \log \tilde{\mathcal L}}{\partial\btheta \partial(\bmath{\delta m})} = \frac{\mathbfss J^{\rm T} \mathbfss V^{-1}}{\gamma},
\label{mM}
\end{equation}
The Hessian $\partial^2\log \tilde{\mathcal L}/\partial\bxi^2$ is usually approximated when
solving the optimization task~(\ref{llopt}). Its structure is similar to the
negative-$\mathbfss F_\bxi$ matrix, for example the $\btheta$-block can be approximated by
$-\mathbfss Q/\gamma$. Therefore, plugging~(\ref{mQ}) and~(\ref{mM}) into~(\ref{Jp}), we
have
\begin{equation}
\frac{\partial\hat\btheta}{\partial(\bmath{\delta m})} \simeq \mathbfss Q^{-1} \mathbfss J^{\rm T} \mathbfss V^{-1},
\end{equation}
while from~(\ref{pvar}) we obtain
\begin{equation}
\var(\delta\hat\btheta) \simeq h \bm{\mathsf\Theta}_2, \quad \bm{\mathsf\Theta}_2 = \mathbfss Q^{-1} \mathbfss J^{\rm T} \mathbfss V^{-1}\mathbfss K_{\rm pert} \mathbfss V^{-1} \mathbfss J \mathbfss Q^{-1}.
\label{Theta2}
\end{equation}
This formula approximates the submatrix of $\bm{\mathsf\Xi}_2$ that corresponds to
$\btheta$.

Concerning the other quantities from~(\ref{pbias},\ref{pvar}), we use Monte Carlo
simulations to assess them. Given the basic best fit of our lightcurves set, we compute the
matrix $\mathbfss K_{\rm pert}$ and then generate Gaussian noise $\bmath{\delta m}$ with
zero mean and this covariance matrix. This simulated perturbation is scaled by $\sqrt{h^*}$
with some small \emph{a priori} selected $h^*$, and then added to the input data which are
then refit. Thus we derive a perturbed trial of $\hat\bxi$ and $\hat{\bm{\mathsf\Xi}}$,
yielding the shifts $\delta\hat\bxi$ and $\delta\hat{\bm{\mathsf\Xi}}$. These shifts are
averaged themselves or in pairwise products necessary to estimate all the
quantities~(\ref{pbias},\ref{pvar}). Finally, the results are divided by $h^*$ to extract
the first-order factors. This yields the estimations for all five required entities: vector
$\bxi_{\rm bias}$, matrices $\bm{\mathsf\Xi}_{\rm bias}$ and $\bm{\mathsf\Xi}_2$, tensors
$\mathsf\Xi_{3,4}$. The alternative analytic approximation~(\ref{Theta2}) can be used for
an additional validation of $\bm{\mathsf\Theta}_2$, the submatrix of $\bm{\mathsf\Xi}_2$.

\section{Interpreting the TTV noise of HD~189733}
\label{sec_HD189733}
Our goal in this section is to fit the transit timing data $\btau$ with some simple model
$\bmu(\bmath p)$, e.g. a linear trend with coefficients $\bmath p$, and simultaneously via
the noise parameter $h$. This will give us the understanding whether the TTV noise excess
of HD~189733 can be modelled through our PBF model, and how physical this model is in
relation with actual observations.

In our analysis we used the same $42$ transit lightcurves selected in
Sect.~\ref{sec_ttvjit}. They included $\sim 7000$ photometric measurements, and so the PBF
perturbation is characterized by the covariance matrix $\mathbfss K_{\rm pert}$ of $\sim
7000\times 7000$ elements. Because of such a size, it cannot be presented here even in
graphical form, but in the online-only material we supply a EPS file showing $\mathbfss
K_{\rm pert}$ as a 2D diagram. Correlations seen in this matrix appear either inside a
single lightcurve, or between lightcurves that refer to the same transit.\footnote{We
assumed here that the perturbation signal is identically the same in such lightcurves. In
actuality it may somewhat differ because of different bandpasses, but in this study we
omitted possible effects of the bandpass dependence.}

From now on, we deal with only a subsample of the fit parameters $\bxi$ that refer to the
midtimes $\btau$. A best fit results in the estimate $\hat\btau$ and its covariance matrix
estimate $\hat{\mathbfss T}$ (which is sampled from $\hat{\bm{\mathsf\Xi}}$). Analogously
to $\hat\btau$ and $\hat{\mathbfss T}$, we can sample subsets from the vector $\bxi_{\rm
bias}$, the matrices $\bm{\mathsf\Xi}_{\rm bias}$, $\bm{\mathsf\Xi}_2$, and the tensors
$\mathsf\Xi_{3,4}$, resulting in some $\btau_{\rm bias}$, $\mathbfss T_{\rm bias}$,
$\mathbfss T_2$, and $\mathsf T_{3,4}$. We computed all these five entities for HD~189733
based on $h^*=10^{-6}$ and $2.5\times 10^5$ Monte Carlo trials. With so big number of
trials, all Monte Carlo uncertainties were negligible.

Let us first consider some preliminary details of how this model works with HD~189733 data.
This is necessary to better understand what should be taken into account, and what can be
neglected.

\begin{figure*}
\includegraphics[width=\linewidth]{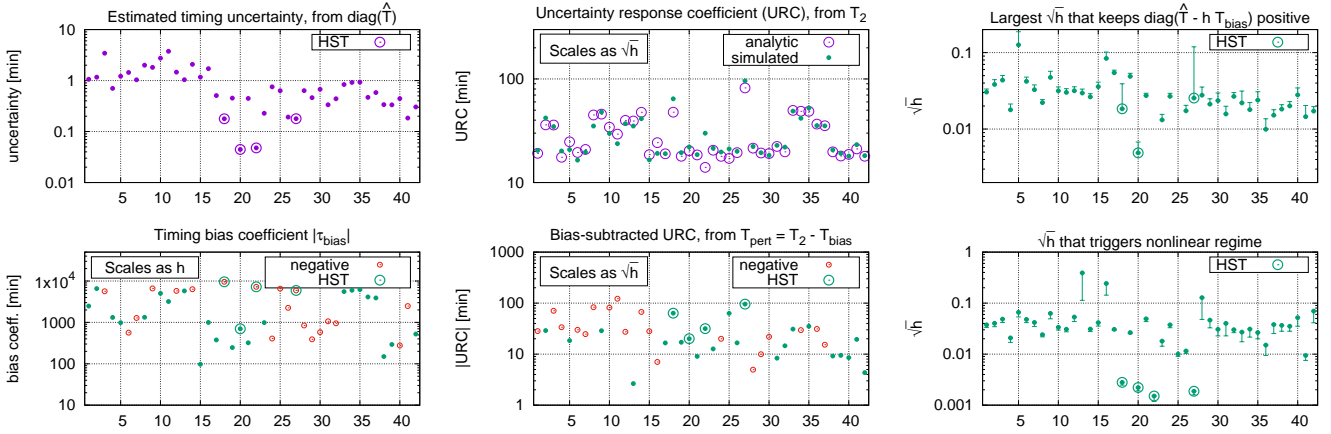}
\caption{Several timing-related characteristics implied by the PBF model of HD~189733. See
text for the explanaition.}
\label{fig_var}
\end{figure*}

First of all, in the left-top panel of Fig.~\ref{fig_var} we can see that timing
measurements are very inhomogeneous regarding their uncertainties. Their accuracy vary from
a few seconds to a few minutes, and this is why analysing such data is so difficult. The
high-accuracy points contribute the largest information in a TTV fit, but simultaneously
they should be more sensitive to various systematic perturbations. Here and in further
plots we especially label four HST timings, because in \citep{Baluevetal21} it was noticed that
these lightcurves do demonstrate some residual trends of unclear nature.

The meaning of the vector $\btau_{\rm bias}$ is easy. The PBF perturbation triggers a
systematic bias in each transit timing, and these biases are approximated by $h \btau_{\rm
bias}$. Notice, however, that we did not expect a bias of this type in
Sect.~\ref{sec_ttvjit}. Basically, our PBF model reveals that TTV noise excess can be
related with timing biases in a mathematically predictable way. The numeric values of
individual $\btau_{\rm bias}$ elements for HD~189733 are shown in left-bottom panel of
Fig.~\ref{fig_var}.

However, our primary effect is the increase of the apparent TTV noise through the matrix
$\mathbfss T_2$. The most important information is carried in its diagonal elements that
describe how the timing variances are affected. To characterize this effect, we introduce
the following Uncertainty Response Coefficients as
\begin{equation}
\text{URC}_i = \sqrt{(T_2)_{ii}}
\label{URCdef}
\end{equation}
The quantity $\text{URC}\times \sqrt h$ has the meaning of the TTV ``jitter'' analogous to
$\sigma_{\rm jit}$ from Sect.~\ref{sec_ttvjit}. However, now this jitter is not constant
and depends on the timing.

Matrix $\mathbfss T_2$ describes how the \emph{true} covariance matrix is perturbed. But in
actuality we deal with the estimate $\hat{\mathbfss T}$ which is biased itself by $h
\mathbfss T_{\rm bias}$. Hence this bias should be subtracted and the correction matrix
$\mathbfss T_2$ should be replaced with
\begin{equation}
\mathbfss T_{\rm pert} = \mathbfss T_2-\mathbfss T_{\rm bias}.
\end{equation}
The URC definition~(\ref{URCdef}) should be transformed to
\begin{equation}
\text{URC}_i' = \sqrt{\left|(T_{\rm pert})_{ii}\right|}.
\end{equation}
This quantity describes the ``effective jitter'' to be added to the \emph{estimated} timing
uncertainty that we actually deal with.

The values of $\text{URC}$ and $\text{URC}'$ for HD~189733 are shown in middle column of
Fig.~\ref{fig_var}. We can see that these coefficients span rather wide numeric range,
about two orders of magnitude, so the additive noise model with a constant jitter could not
model the PBF effect correctly. Notice that $\mathbfss T_2$ is positive definite, by
definition, so its diagonal elements could be only positive. But $\mathbfss T_{\rm pert}$
can be indefinite in general and may include positive as well as negative diagonal
elements. That is, the PBF effect may systematically increase or systematically decrease
the timing uncertainty. Negative occurences mean that the corresponding uncertainty is
expectedly overvalued because of the perturbation.

\begin{figure*}
\includegraphics[width=\linewidth]{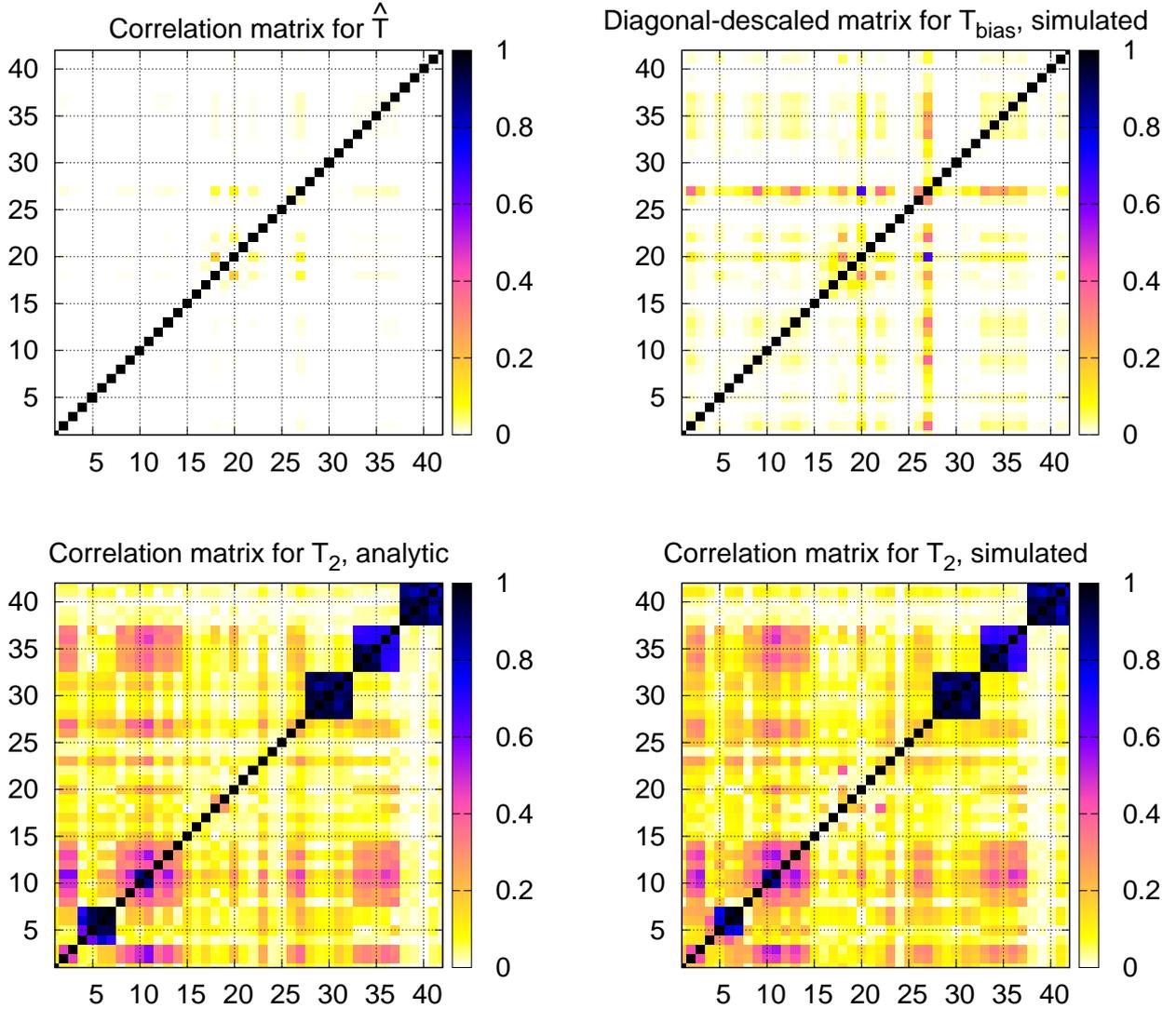}
\caption{Correlation matrices implied by the PBF model of HD~189733. See text for the
explanaition.}
\label{fig_covars}
\end{figure*}

In Fig.~\ref{fig_covars} we also plot 2D diagrams related to several correlation matrices:
the correlation matrix for $\tau_i$, the correlation matrix corresponding to $\mathbfss
T_2$, for its analytic and simulated approximations, and the diagonal-descaled
(correlation-like) matrix for $\mathbfss T_{\rm bias}$. We plot only absolute values
disregarding the signs. From these plots we can see that $\mathbfss T_2$ may boost
significant correlations between timings, even though such correlations are not seen in
$\hat{\mathbfss T}$. The effect from $\mathbfss T_{\rm bias}$ on timing correlations is
smaller. Also, from both Fig.~\ref{fig_var} (middle-top panel) and Fig.~\ref{fig_covars}
(two bottom panels) it follows that analytic and simulated versions of $\mathbfss T_2$ are
in good agreement with each other.

The further effect on timing uncertainties comes from the tensors $\mathsf T_{3,4}$ that
describe random errors in $\hat{\mathbfss T}$ owed to the PBF. However, their practical
impact appeared more complicated, and before considering it we need to understand the
applicability ranges for our TTV model. In view of that two additional characteristics can
be considered, as defined below.

According to the physical meaning of $\mathbfss T_{\rm bias}$, the difference
$\hat{\mathbfss T} - h \mathbfss T_{\rm bias}$ represents the ``unperturbed'' covariance
matrix of $\btau$, i.e. the one that would appear without PBF perturbation. Therefore,
$\hat{\mathbfss T} - h \mathbfss T_{\rm bias}$ should necessarily be a positive definite
matrix. In particular all its diagonal elements must be positive, otherwise such a model
appears nonphysical. This restriction allows to derive an upper limit on $h$. This limit,
however, can be somewhat relaxed because each $(\hat T)_{ii}$ has a probable error about
$\sqrt{h (T_4)_{iiii}}$. Assuming that $(\hat T)_{ii}$ is undervalued by $s \sqrt{h
(T_4)_{iiii}}$ with some factor $s$ (number of sigma), the ``unperturbed'' variance should
then be $(\hat T)_{ii} + s \sqrt{h (T_4)_{iiii}} - h (T_{\rm bias})_{ii}$. Requiring it to
be positive results in an inequality that can be solved for $\sqrt h$, yielding an
$s$-sigma upper boundary:
\begin{eqnarray}
(\hat T)_{ii} + s \sqrt{h (T_4)_{iiii}} - h (T_{\rm bias})_{ii}\geq 0 \implies \nonumber\\
\sqrt h \leq \left\{ \begin{array}{ll} \frac{\sqrt{(T_4)_{iiii}} + \sqrt{(T_4)_{iiii} + 4(\hat T)_{ii} (T_{\rm bias})_{ii}}}{2(T_{\rm bias})_{ii}}, & s=1, \\
\sqrt{\frac{(\hat T)_{ii}}{(T_{\rm bias})_{ii}}}, & s=0. \end{array}\right.
\label{hphys}
\end{eqnarray}
This requres $(T_{\rm bias})_{ii}>0$. If it turns negative for some $i$, the corresponding
diagonal element in $\hat{\mathbfss T} - h \mathbfss T_{\rm bias}$ keeps positive for any
$h>0$, and hence such transits set no limit on $h$.

In Fig.~\ref{fig_var}, right-top panel, we plot the boundaries~(\ref{hphys}) as points with
one-sided errorbars. We can see that most of them reside above the level $h=0.01$, except
for a single HST observation. Therefore, $h\gtrsim 0.01$ or so renders our PBF model
definitely non-physical (though it may remain formally tractable in mathematical sense).
The range from $h\sim 0.005$ to $h \sim 0.01$ is somewhat disputable, because we have only
a single point there. This may indicate, alternatively, that this particular HST lightcurve
is odd in some sense (e.g. it may have an undervalued uncertainty owed to some overfit
effect).

Another model limitation appears because we considered only linear effects in terms of $h$.
This necessitates that all corrections to $\hat{\mathbfss T}$ should remain small. The
systematic correction to $(\hat T)_{ii}$ is $h (T_{\rm pert})_{ii}$, while the correction
due to random errors is $\pm s\sqrt{h (T_4)_{iii}}$. The maximum absolute correction should
be kept small in comparison with $(\hat T)_{ii}$, resulting in the following limits:
\begin{eqnarray}
h \left|(T_{\rm pert})_{ii}\right| + s \sqrt{h (T_4)_{iiii}} \lesssim (\hat T)_{ii} \implies \nonumber\\
\sqrt h \gtrsim \left\{ \begin{array}{ll} \frac{ \sqrt{(T_4)_{iiii} + 4(\hat T)_{ii} \left|(T_{\rm pert})_{ii}\right|} - \sqrt{(T_4)_{iiii}}}{2\left|(T_{\rm pert})_{ii}\right|}, & s=1, \\
\sqrt{\frac{(\hat T)_{ii}}{\left|(T_{\rm pert})_{ii}\right|}}, & s=0. \end{array}\right.
\label{hlin}
\end{eqnarray}

The boundaries~(\ref{hlin}) are shown in the right-bottom panel of Fig.~\ref{fig_var},
again with one-sided errorbars. To keep our model mathematically accurate, $h$ should stay
below them. Larger $h$ does not turn the model turn non-physical, but our linear
approximation may appear inaccurate without higher-order terms in $h$. We can see that
majority of the data imply the limit $h\lesssim 0.01$, just like previously. However, four
HST points reside in the range $h \sim 0.001$ to $h\sim 0.003$. It is not entirely clear,
whether these four points may corrupt our TTV analysis, but this range of $h$ should be
considered with care again.

\begin{figure}
\includegraphics[width=84mm]{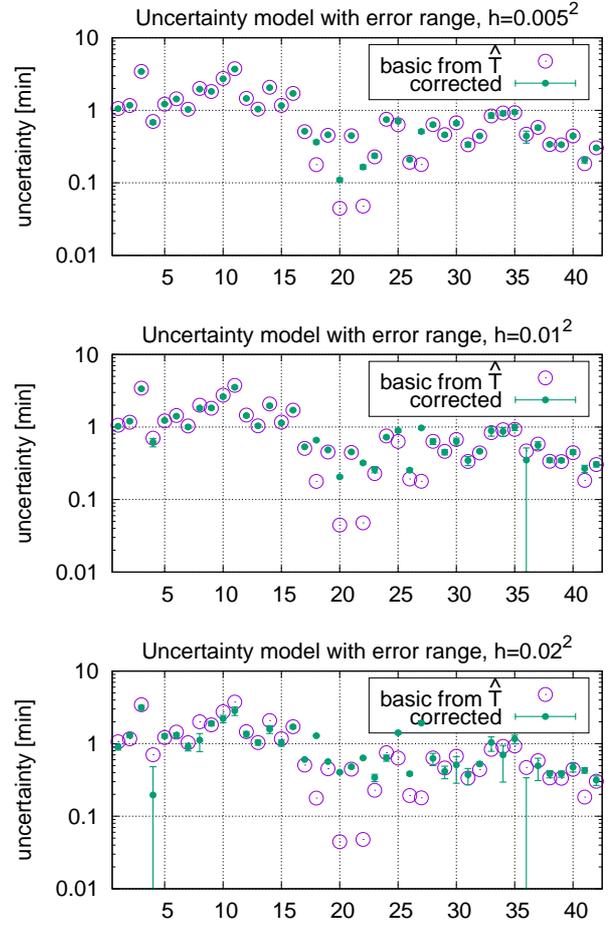}
\caption{Corrected timing uncertainties implied by the PBF model of HD~189733, as derived
from three reference values of the $h$ parameter. Each uncertainty includes a error bar
owed to the statistical nature of the PBF perturbation.}
\label{fig_vmod}
\end{figure}

Now let us return to the effect of random errors in $\hat{\mathbfss T}$. They can make our
task considerably more complicated from the analytic point of view, because the analysis
should include some treatment regarding the ``uncertainty of the uncertainty'' phenomenon.
This change may cause an increase of the complexity, but currently it is unclear how big
this effect is and whether it should be taken into account. Now, when reasonable range for
$h$ is established, the easiest way to assess this effect is to compute it for practical
data and a few sample $h$. This is presented in Fig.~\ref{fig_vmod}, where we show
corrected timing uncertainties (from $\hat{\mathbfss T}+h\mathbfss T_{\rm pert}$) and their
potential errorbars (from $h \mathsf T_4$). We can see that nearly all errorbars remein
negligible even for $h=0.02$. For $h=0.01$ just a single transit demonstrates a big
errorbar, and for $h=0.005$ all errorbars are negligible. This enables us to conclude that
the total effect from the covariance tensors $\mathsf T_{3,4}$ is likely not important in
the context of these particular data of HD~189733. We can analyse these data using a more
traditional approach with deterministic uncertainties.

To fit the TTV data honouring the remaining (non-random) corrections, we should subtract
the bias $h\btau_{\rm bias}$ from $\hat\btau$, and correct the covariance matrix
$\hat{\mathbfss T}$ by adding $h \mathbfss T_{\rm pert}$. Assuming that timing errors are
Gaussian, the likelihood function becomes similar to~(\ref{ll}):
\begin{eqnarray}
\log \mathcal L_{\rm TTV}(\bmath p, h) = -\frac{1}{2} \log \det\mathbfss T - \frac{1}{2\gamma} \bmath r^{\rm T} \mathbfss T^{-1} \bmath r, \nonumber\\
\mathbfss T = \hat{\mathbfss T}+h\mathbfss T_{\rm pert}, \quad \bmath r = \bmu(\bmath p) - \hat\btau + h\btau_{\rm bias}, \nonumber\\ \gamma=1-\frac{\dim \bmath p}{N_\btau}.
\label{llttv}
\end{eqnarray}
The best fitting estimate is obtained by maximizing this $\mathcal L_{\rm TTV}$ with
respect to $h$ and $\bmath p$. Notice that $h$ is a mixed-type parameter, affecting both
the TTV noise (through $\mathbfss T$) and the TTV curve (through $\bmath r$).

\begin{table*}
\caption{TTV fits of the PBF model for HD~189733.}
\label{tab_PBFfits}
\begin{tabular}{llll}
\hline
TTV model $\bmu(\bmath p)$ & Goodness-of-fit $\tilde l$ [s] & $\varkappa=\frac{\sqrt h}{1.086}$ & Periodogram peak power \\
\hline
\multicolumn{3}{c}{Initial set of $42$ lightcurves:}\\
Linear    & $349.84$ & $0.0090\pm 0.0012$  & $\mathbf{80.4}$ \\
Quadratic & $215.77$ & $0.0088\pm 0.0013$  & $\mathbf{62.1}$ \\
\multicolumn{3}{c}{minus $4$ HST lightcurves:}\\
Linear    & $436.70$ & $0.0092\pm 0.0016$  & $\mathbf{78.0}$ \\
Quadratic & $150.53$ & $0.0066\pm 0.0023$  & $\mathbf{45.4}$ \\
\hline
\end{tabular}
\end{table*}
The results of these TTV fits for HD~189733 are presented in Table~\ref{tab_PBFfits}. We
use two models for $\bmu(\bmath p)$: linear or quadratic trend. Our primary fits involve
$42$ timings selected in Sect.~\ref{sec_ttvjit}. In the context of the above-discussed TTV
effects, \citet{Kasper19} data do not reveal any odd behaviour, however the four HST
timings do: they have especially low formal uncertainty, hence highest sensitivity to the
PBF perturbation, and so inspire higher model nonlinearity. Besides, as noticed above, they
may involve incompletely reduced photometric trends that may or may not be related to the
hypothetical PBF effect. Therefore, we performed additional analysis for only $38$ timings
removing the HST ones.

From these results, we can make the following conclusions:
\begin{enumerate}
\item The goodness-of-fit measure of the new fits appears significantly worse than in
Sect.~\ref{sec_ttvjit}. This suggests that PBF model can incorporate only a fraction of the
observed TTV excess.

\item The best fitting $\varkappa$ parameter of the PBF model usually appears near its
upper physical bound, that is the model tries to select the largest $\varkappa$ still
admissible. This again means that the model is largerly deficient in explaining the TTV
data.

\item The formally fitted $\varkappa$ corresponds, most closely, to the middle panel of
Fig.~\ref{fig_vmod}. From this plot we can see that PBF perturbation mainly affects just
the four timings coming from HST, and, possibly, a couple of others. Their initial formal
uncertainties can be below $3$~s, but should be boosted to, at least, $\sim 10$~s. From one
side, this again confirms that HST lightcurves contain some perturbing trends, and our PBF
model can make their uncertainties more adequate. However, the previously estimated TTV
jitter from Table~\ref{tab_ttvfits} appears much larger than $10$~s, suggesting that we
should also correct mid-accuracy timings, at least.

\item A `forced' increase of $\varkappa$ too much above the $\sim 0.01$ level is impossible
because some timing variances turn negative. This level defines the natural applicability
limit for our PBF model and its maximum capabilities in how much it can explain the TTV
jitter.
\end{enumerate}

\begin{figure*}
\includegraphics[width=\linewidth]{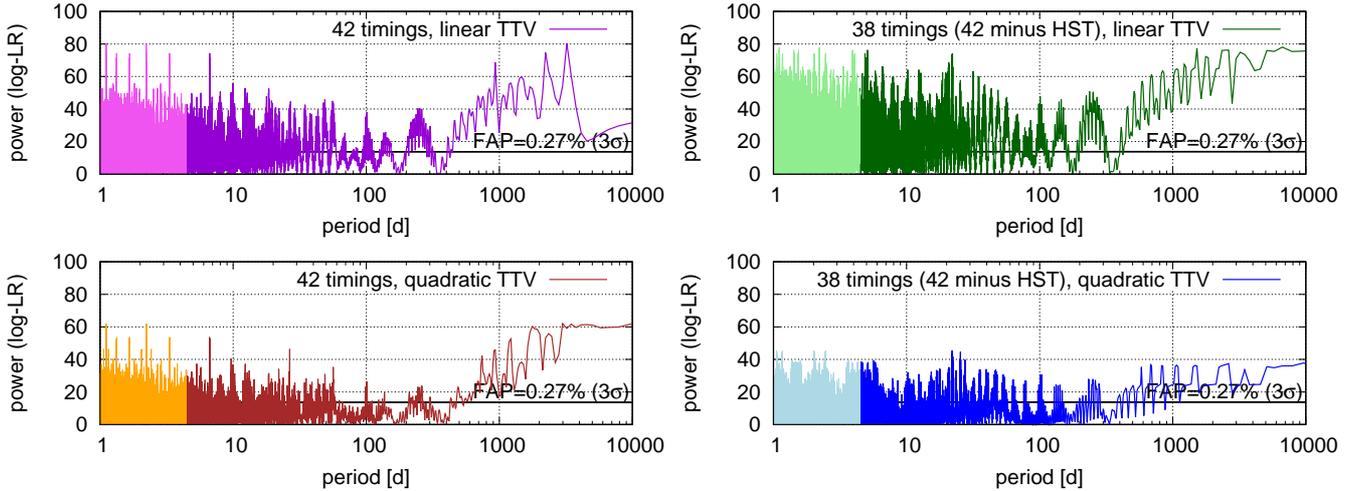}
\caption{Likelihood-based periodograms of HD~189733 timings,
for the PBF fits from Table~\ref{tab_PBFfits}. Details are similar to Fig.~\ref{fig_prdg}.}
\label{fig_prdg_PBF}
\end{figure*}

Because our PBF treatment keeps most of the uncertainties nearly intact, the formal
significance of various putative TTV signals is greatly increased. For example, models with
a quadratic TTV trend now have drastically larger likelihood than those with the linear one
(they did not differ signfificantly in Sect.~\ref{sec_ttvjit}). The corresponding residual
periodograms (see Fig.~\ref{fig_prdg_PBF}) also exceed formal significance level. It is
still inobvious how trustable these periodograms can be, as their look is very
model-dependent. But at least, from Table~\ref{tab_PBFfits} it follows that by making the
TTV signal more complicated the remaining TTV excess can be significantly reduced.

Therefore, the main conclusion from our analysis is that random PBF cannot explain the
observed TTV jitter well. The largest admissible $\varkappa$ can incorporate only a
fraction of the observed TTV jitter, while further increase of $\varkappa$ makes the model
meaningless instead of generating a higher TTV noise.

On the other hand, our results do not reject the PBF effect for HD~189733. It can explain
some part of the TTV jitter, at the maximum level of $\sim 10$~s. This is an important
observational constraint on a physical effect that cannot be directly observed. In such a
case, our estimation $\varkappa\simeq 0.01$ should be considered as an upper limit. If
$\varkappa$ was larger, for some transits we could not obtain so small timing uncertainties
as we have, because their low limit is set by $h \mathbfss T_{\rm bias}$. Notice that such
$\varkappa$ would correspond to the third plot of Fig.~\ref{fig_BF}, for example.

\section{Discussion}
In this work we designed a method to treat the impact of stellar photosperic field on
exoplanetary transits. Our method does not assume any specific physical nature of
brightness inhomogeneities and solves the task in a nearly model-invariable way. Namely,
the PBF effect on the transit signal is proportional to just a single aggregate
characteristic $\varkappa$, basically the PBF power spectrum taken at zero argument,
$P_I(0)$. The correlation function of the photometric perturbation is universal and
depends, in a mathematically defined way, only on the usual transit parameters and limb
darkening. Therefore, the PBF perturbation can be fitted as a random process, through its
scaling parameter $\varkappa$ that has direct physical meaning.

Summarizing all the investigations regarding HD~189733, the answer to our primary question,
whether any photosperic brightness inhomogeneities can explain its TTV excess, is likely
negative. The PBF effect cannot boost timings uncertainties consistently by more than $\sim
10$~s, while the actual TTV excess is $\sim 70$~s.

Therefore, the detailed source of TTV noise of HD~189733 remains unclear. It seems likely
that there are multiple sources of this TTV noise that interfere in a complicated way. As
discussed in Sect.~\ref{sec_ttvjit}, possible pipeline inaccurcies may undervalue derived
timing uncertainties by $\sim 20\%$, but this again leaves most of TTV noise unexplained.
It seems we should consider more seriously that some part of the TTV has a deterministic
nature, e.g. a perturbation from another gravitating body. But because of complicated noise
contamination we cannot characterize it well. Other possible sources that yet need a
detailed investigation:
\begin{enumerate}
\item The effect of planetary atmosphere.
\item Gravitational effect from hypothetical planets in a chaotic (but stable) dynamics, so
that their TTV signal looks like noise.
\item Possible more complicated time-variable activity effects, e.g. related to the star
magnetic activity cycle.
\item Incompletely compensated instrumental photometric drifts that occasionally
contaminated some high-accuracy lightcurves.
\item Some lightcurves may still involve erratic BJD/HJD correction (such cases do appear
even in refereed papers, see Table~1 in \citealt{Baluevetal21}).
\end{enumerate}

Returning to the PBF effect and its modelling method that we developed, it is probably
interesting to verify them with other planet hosts considered in \citep{Baluev19}. Some of
them also had a statistically significant TTV excess, though with a smaller magnitude than
for HD~189733. The PBF model may appear more successful in such cases. However, the primary
practical trouble with such analysis is the need of Monte Carlo simulations. Simulations
are necessary to estimate all required corrections, in particular $\mathbfss T_{\rm bias}$
and $\btau_{\rm bias}$. Our method involves analytic compution of $\mathbfss T_2$, and it
demonstrated good agreement with simulations, but $\mathbfss T_{\rm bias}$ and $\btau_{\rm
bias}$ appeared too difficult for the analytic treatment. Resolving this issue can make the
entire analysis pipeline drasticaly faster and more practical for a routine use.

\section*{Acknowledgments}
This work was supported by the Russian Science Foundation, project 19-72-10023. We highly
appreciate fruitful comments and suggestions provided by the anonymous reviewer.

\section*{Data availability}
The data underlying this article are available in the article and in its online
supplementary material.

\bibliographystyle{mnras}
\bibliography{transitnoise}

\begin{thebibliography}{}
\makeatletter
\relax
\def\mn@urlcharsother{\let\do\@makeother \do\$\do\&\do\#\do\^\do\_\do\%\do\~}
\def\mn@doi{\begingroup\mn@urlcharsother \@ifnextchar [ {\mn@doi@}
  {\mn@doi@[]}}
\def\mn@doi@[#1]#2{\def\@tempa{#1}\ifx\@tempa\@empty \href
  {http://dx.doi.org/#2} {doi:#2}\else \href {http://dx.doi.org/#2} {#1}\fi
  \endgroup}
\def\mn@eprint#1#2{\mn@eprint@#1:#2::\@nil}
\def\mn@eprint@arXiv#1{\href {http://arxiv.org/abs/#1} {{\tt arXiv:#1}}}
\def\mn@eprint@dblp#1{\href {http://dblp.uni-trier.de/rec/bibtex/#1.xml}
  {dblp:#1}}
\def\mn@eprint@#1:#2:#3:#4\@nil{\def\@tempa {#1}\def\@tempb {#2}\def\@tempc
  {#3}\ifx \@tempc \@empty \let \@tempc \@tempb \let \@tempb \@tempa \fi \ifx
  \@tempb \@empty \def\@tempb {arXiv}\fi \@ifundefined
  {mn@eprint@\@tempb}{\@tempb:\@tempc}{\expandafter \expandafter \csname
  mn@eprint@\@tempb\endcsname \expandafter{\@tempc}}}

\bibitem[\protect\citeauthoryear{Abubekerov \& Gostev}{Abubekerov \&
  Gostev}{2013}]{AbubGost13}
Abubekerov M.~K.,  Gostev N.~Y.,  2013, \mnras, 432, 2216

\bibitem[\protect\citeauthoryear{Agol, Steffen, Sari  \& Clarkson}{Agol
  et~al.}{2005}]{Agol05}
Agol E.,  Steffen J.,  Sari R.,   Clarkson W.,  2005, \mnras, 359, 567

\bibitem[\protect\citeauthoryear{Bailey \& Goodman}{Bailey \&
  Goodman}{2019}]{BaileyGoodman19}
Bailey A.,  Goodman J.,  2019, \mnras, 482, 1872

\bibitem[\protect\citeauthoryear{Bakos et~al.,}{Bakos et~al.}{2006}]{Bakos06}
Bakos G.~A.,  et~al., 2006, \apj, 650, 1160

\bibitem[\protect\citeauthoryear{Baluev}{Baluev}{2008}]{Baluev08a}
Baluev R.~V.,  2008, \mnras, 385, 1279

\bibitem[\protect\citeauthoryear{Baluev}{Baluev}{2009}]{Baluev08b}
Baluev R.~V.,  2009, \mnras, 393, 969

\bibitem[\protect\citeauthoryear{Baluev}{Baluev}{2012}]{Baluev12}
Baluev R.~V.,  2012, \mnras, 422, 2372

\bibitem[\protect\citeauthoryear{Baluev}{Baluev}{2013a}]{Baluev13c}
Baluev R.~V.,  2013a, \ac, 2, 18

\bibitem[\protect\citeauthoryear{Baluev}{Baluev}{2013b}]{Baluev13a}
Baluev R.~V.,  2013b, \mnras, 429, 2052

\bibitem[\protect\citeauthoryear{Baluev}{Baluev}{2015}]{Baluev14a}
Baluev R.~V.,  2015, \mnras, 446, 1493

\bibitem[\protect\citeauthoryear{Baluev}{Baluev}{2018}]{Baluev18c}
Baluev R.~V.,  2018, \ac, 25, 221

\bibitem[\protect\citeauthoryear{Baluev \& Shaidulin}{Baluev \&
  Shaidulin}{2015}]{BaluevShaidulin15}
Baluev R.~V.,  Shaidulin V.~S.,  2015, \mnras, 454, 4379

\bibitem[\protect\citeauthoryear{Baluev et~al.,}{Baluev
  et~al.}{2015}]{Baluev15a}
Baluev R.~V.,  et~al., 2015, \mnras, 450, 3101

\bibitem[\protect\citeauthoryear{Baluev et~al.,}{Baluev
  et~al.}{2019}]{Baluev19}
Baluev R.~V.,  et~al., 2019, \mnras, 490, 1294

\bibitem[\protect\citeauthoryear{Baluev et~al.,}{Baluev
  et~al.}{2020}]{Baluev20}
Baluev R.~V.,  et~al., 2020, \mnras, 496, L11

\bibitem[\protect\citeauthoryear{Baluev et~al.,}{Baluev
  et~al.}{2021}]{Baluevetal21}
Baluev R.~V.,  et~al., 2021, Acta Astron., 71, 25

\bibitem[\protect\citeauthoryear{Boisse et~al.,}{Boisse
  et~al.}{2009}]{Boisse09}
Boisse I.,  et~al., 2009, \aap, 495, 959

\bibitem[\protect\citeauthoryear{Bouma et~al.,}{Bouma et~al.}{2019}]{Bouma19}
Bouma L.~G.,  et~al., 2019, \aj, 157, 217

\bibitem[\protect\citeauthoryear{Bouma, Winn, Howard, Howell, Isaacson, Knutson
   \& Matson}{Bouma et~al.}{2020}]{Bouma20}
Bouma L.~G.,  Winn J.~N.,  Howard A.~W.,  Howell S.~B.,  Isaacson H.,  Knutson
  H.,   Matson R.~A.,  2020, \apj, 893, L29

\bibitem[\protect\citeauthoryear{Cauley, Redfield  \& Jensen}{Cauley
  et~al.}{2017}]{Cauley17}
Cauley P.~W.,  Redfield S.,   Jensen A.~G.,  2017, \aj, 153, 217

\bibitem[\protect\citeauthoryear{Chiavassa et~al.,}{Chiavassa
  et~al.}{2017}]{Chiavassa17}
Chiavassa A.,  et~al., 2017, \aap, 597, A94

\bibitem[\protect\citeauthoryear{Claret \& Bloemen}{Claret \&
  Bloemen}{2011}]{ClaretBloemen11}
Claret A.,  Bloemen S.,  2011, \aap, 529, A75

\bibitem[\protect\citeauthoryear{Ford et~al.,}{Ford et~al.}{2011}]{Ford11}
Ford E.~B.,  et~al., 2011, \aap, 197, 2

\bibitem[\protect\citeauthoryear{Holman \& Murray}{Holman \&
  Murray}{2005}]{HolmanMurray05}
Holman M.~J.,  Murray N.~W.,  2005, Science, 307, 1288

\bibitem[\protect\citeauthoryear{Karpinskii \& Mekhanikov}{Karpinskii \&
  Mekhanikov}{1977}]{Karpinskii77}
Karpinskii V.~N.,  Mekhanikov V.~V.,  1977, Solar Physics, 54, 25

\bibitem[\protect\citeauthoryear{Kasper et~al.,}{Kasper
  et~al.}{2019}]{Kasper19}
Kasper D.~H.,  et~al., 2019, \mnras, 483, 3781

\bibitem[\protect\citeauthoryear{Maciejewski et~al.,}{Maciejewski
  et~al.}{2016}]{Maciejewski16}
Maciejewski G.,  et~al., 2016, \aap, 588, L6

\bibitem[\protect\citeauthoryear{Maciejewski et~al.,}{Maciejewski
  et~al.}{2018}]{Maciejewski18a}
Maciejewski G.,  et~al., 2018, Acta Astronomica, 68, 371

\bibitem[\protect\citeauthoryear{McCullough, Crouzet, Deming  \&
  Madhusudhan}{McCullough et~al.}{2014}]{McCullough14}
McCullough P.~R.,  Crouzet N.,  Deming D.,   Madhusudhan N.,  2014, \apj, 791,
  55

\bibitem[\protect\citeauthoryear{Montalto, Bou{\'e}, Oshagh, Boisse, Bruno  \&
  Santos}{Montalto et~al.}{2014}]{Montalto14}
Montalto M.,  Bou{\'e} G.,  Oshagh M.,  Boisse I.,  Bruno G.,   Santos N.~C.,
  2014, \mnras, 444, 1721

\bibitem[\protect\citeauthoryear{Nesis, Hammer, Roth  \& Schleicher}{Nesis
  et~al.}{2002}]{Nesis02}
Nesis A.,  Hammer R.,  Roth M.,   Schleicher H.,  2002, \aap, 396, 1003

\bibitem[\protect\citeauthoryear{Pillitteri, Wolk, Lopez-Santiago, G{\"u}nther,
  Sciortino, Cohen, Kashyap  \& Drake}{Pillitteri et~al.}{2014}]{Pillitteri14}
Pillitteri I.,  Wolk S.~J.,  Lopez-Santiago J.,  G{\"u}nther H.~M.,  Sciortino
  S.,  Cohen O.,  Kashyap V.,   Drake J.~J.,  2014, \apj, 785, 145

\bibitem[\protect\citeauthoryear{Pont et~al.,}{Pont et~al.}{2007}]{Pont07}
Pont F.,  et~al., 2007, \aap, 476, 1347

\bibitem[\protect\citeauthoryear{Santos et~al.,}{Santos
  et~al.}{2013}]{Santos13}
Santos N.~C.,  et~al., 2013, \aap, 556, A150

\bibitem[\protect\citeauthoryear{Sokov et~al.,}{Sokov et~al.}{2018}]{Sokov18}
Sokov E.~N.,  et~al., 2018, \mnras, 480, 291

\bibitem[\protect\citeauthoryear{Steffen et~al.,}{Steffen
  et~al.}{2013}]{Steffen13}
Steffen J.~H.,  et~al., 2013, \mnras, 428, 1077

\bibitem[\protect\citeauthoryear{Vuong}{Vuong}{1989}]{Vuong89}
Vuong Q.~H.,  1989, Econometrica, 57, 307

\bibitem[\protect\citeauthoryear{Winn et~al.,}{Winn et~al.}{2007}]{Winn07a}
Winn J.~N.,  et~al., 2007, \aj, 133, 1828

\bibitem[\protect\citeauthoryear{Xie}{Xie}{2013}]{Xie13}
Xie J.-W.,  2013, \apjs, 208, 22

\makeatother
\end{thebibliography}

\appendix

\section{Online material}
The paper contains the following online files:
\begin{enumerate}
\item ZIP archive with a C++ library that allows to compute the $k_{\rm pert}$ function.
\item Scaleable EPS figure that shows full correlation matrix $\mathbfss K_{\rm pert}$
computed for $42$ transit lightcurves of HD~189733.
\end{enumerate}

\bsp

\label{lastpage}

\end{document}